\documentclass[journal]{IEEEtran}

\ifCLASSINFOpdf

\else

\fi

\usepackage{algorithmic}
\usepackage{cite}
\usepackage{algorithm}
\usepackage{amsmath,amssymb,amsfonts}
\usepackage{algorithmic}
\usepackage{graphicx}
\usepackage{textcomp}
\usepackage{xcolor}
\graphicspath{ {./Figures/} }
\usepackage{epsfig}
\usepackage{epstopdf}
\usepackage{enumitem}
\usepackage{url}
\usepackage{glossaries}
\usepackage{comment, soul}
\usepackage{multirow}

\usepackage{subfig}

\usepackage{todonotes}
\usepackage{tikz}
\usepackage{footmisc}

\usepackage{pifont}
\newcommand{\cmark}{\ding{52}}%
\newcommand{\xmark}{\ding{54}}%

\usepackage{xcolor,colortbl}
\definecolor{Gray}{gray}{0.95}

\usepackage{acronym}
\newacro{hevc}[HEVC]{high efficiency video coding}
\newacro{sai}[SAI]{{\it subaperture-image}}
\newacro{mi}[MI]{{\it micro-image}}
\newacro{lf}[LF]{light field}
\newacro{vr}[VR]{virtual reality}
\newacro{llf}[LLF]{{\it lenslet light field}}

\newacro{vceg}[VCEG]{video coding experts group}
\newacro{mpeg}[MPEG]{motion picture experts group}
\newacro{vvc}[VVC]{versatile video coding}
\newacro{lf-d2gan}[LF-D2GAN]{LF dual discriminator generative adversarial network}
\newacro{kl}[LK]{Kullback-Leibler}
\newacro{mv-qenn}[MV-QENet]{multi-view quality enhancement convolutional neural Network}
\newacro{psnr}[PSNR]{peak signal-to-noise ratio} 
\newacro{dsc}[DSC]{distributed source coding} 
\newacro{dct}[DCT]{discrete cosine transform} 
\newacro{dst}[DST]{discrete sine transform} 
\newacro{bi-ss}[BI-SS]{bi-prediction self-similarity} 
\newacro{cabac}[CABAC]{context-adaptive binary arithmetic coding} 
\newacro{cnn}[CNN]{convolutional neural network} 
\newacro{gan}[GAN]{generative adversarial network} 
\newacro{asr}[ASR]{angular super-resolution} 
\newacro{mstn}[MSTN]{multibranch spatial transformer networks}
\newacro{ssim}[SSIM]{structural similarity index measure}
\newacro{cornia}[CORNIA]{codebook representation for no-reference image assessment}
\newacro{iqa}[IQA]{image quality assessment}
\newacro{mos}[MOS]{mean opinion scores}
\newacro{rvs}[RVS]{reference view selection}
\newacro{qp}[QP]{quantization parameter}
\newacro{si}[SI]{spatial information}
\newacro{cf}[CF]{color fulness}
\newacro{ra}[RA]{random access}
\newacro{vtm}[VTM]{VVC test model}
\newacro{bd-br}[BD-BR]{Bj\o ntegaard delta bit rate}
\newacro{bd-psnr}[BD-\acs{psnr}]{Bj\o ntegaard delta \acs{psnr}}
\newacro{bd-ssim}[BD-\acs{ssim}]{Bj\o ntegaard delta \acs{ssim}}
\newacro{rdo}[RDO]{rate-distortion optimization}
\newacro{rd}[RD]{rate-distortion}
\newacro{dc}[DC]{direct current}
\newacro{ac}[AC]{alternating current}
\newacro{lfnst}[LFNST]{low frequency non-separable transform}
\newacro{sbtmvp}[SbTMVP]{subblock-based temporal motion vector prediction}
\newacro{aff}[AFF]{affine motion model}
\newacro{bdof}[BDOF]{bi-directional optical flow}
\newacro{mv}[MV]{motion vector}
\newacro{dmvr}[DMVR]{decoder-side motion vector refinement}
\newacro{poc}[POC]{picture order count} 
\newacro{gop}[GOP]{groups of pictures} 
\newacro{jem}[JEM]{joint exploration model} 
\newacro{jvet}[JVET]{joint video exploration team} 
\newacro{mrl}[MRL]{multiple reference line} 
\newacro{mip}[MIP]{matrix Intra prediction} 
\newacro{bn}[BN]{batch normalization} 
\usepackage{hyperref}
\begin{document}

\title{Light Field Image Coding Using VVC standard and View Synthesis based on Dual Discriminator GAN}
\author{Nader Bakir,
        Wassim~Hamidouche,~\IEEEmembership{Member,~IEEE},
        Sid~Ahmed~Fezza,
        Khouloud Samrouth,
        and~Olivier~D\'eforges
\thanks{N. Bakir, W. Hamidouche and O. D\'eforges are with Univ. Rennes, INSA Rennes, CNRS, IETR - UMR 6164, Rennes, France (e-mail: \href{mailto:whamidou@insa-rennes.fr}{whamidou@insa-rennes.fr}).}
\thanks{SA. Fezza is with National Institute of Telecommunications and ICT, Oran, Algeria (e-mail: \href{mailto:sfezza@inttic.dz}{sfezza@inttic.dz}).}
\thanks{K. Samrouth is with Lebanese University, Tripoli, Lebanon (e-mail: \href{mailto:khouloud.samrout@gmail.com}{khouloud.samrout@gmail.com}).}
}

\markboth{Accepted Version}
{Bakir \MakeLowercase{\textit{et al.}}: LF Coding using VVC and D2GAN}

\maketitle

\begin{abstract}
Light field (LF) technology is considered as a promising way for providing a high-quality virtual reality (VR) content. However, such an imaging technology produces a large amount of data requiring efficient LF image compression solutions. In this paper, we propose a LF image coding method based on a view synthesis and view quality enhancement techniques. Instead of transmitting all the LF views, only a sparse set of reference views are encoded and transmitted, while the remaining views are synthesized at the decoder side. The transmitted views are encoded using the versatile video coding (VVC) standard and are used as reference views to synthesize the dropped views. The selection of non-reference dropped views is performed using a rate-distortion optimization based on the VVC temporal scalability. The dropped views are reconstructed using the LF dual discriminator GAN (LF-D2GAN) model. In addition, to ensure that the quality of the views is consistent, at the decoder, a quality enhancement procedure  is performed on the reconstructed views allowing smooth navigation across views. Experimental results show that the proposed method provides high coding performance and overcomes the state-of-the-art LF image compression methods by --36.22\% in terms of BD-BR and 1.35 dB in BD-PSNR. The web page of this work is available at \color[rgb]{0.858,0.188,0.478}{\url{https://naderbakir79.github.io/LFD2GAN.html}}.
\end{abstract}


\IEEEpeerreviewmaketitle

\section{Introduction}
\label{sec:intro}

\begin{figure}[t!]
	\centerline{\includegraphics[width=0.4\textwidth]{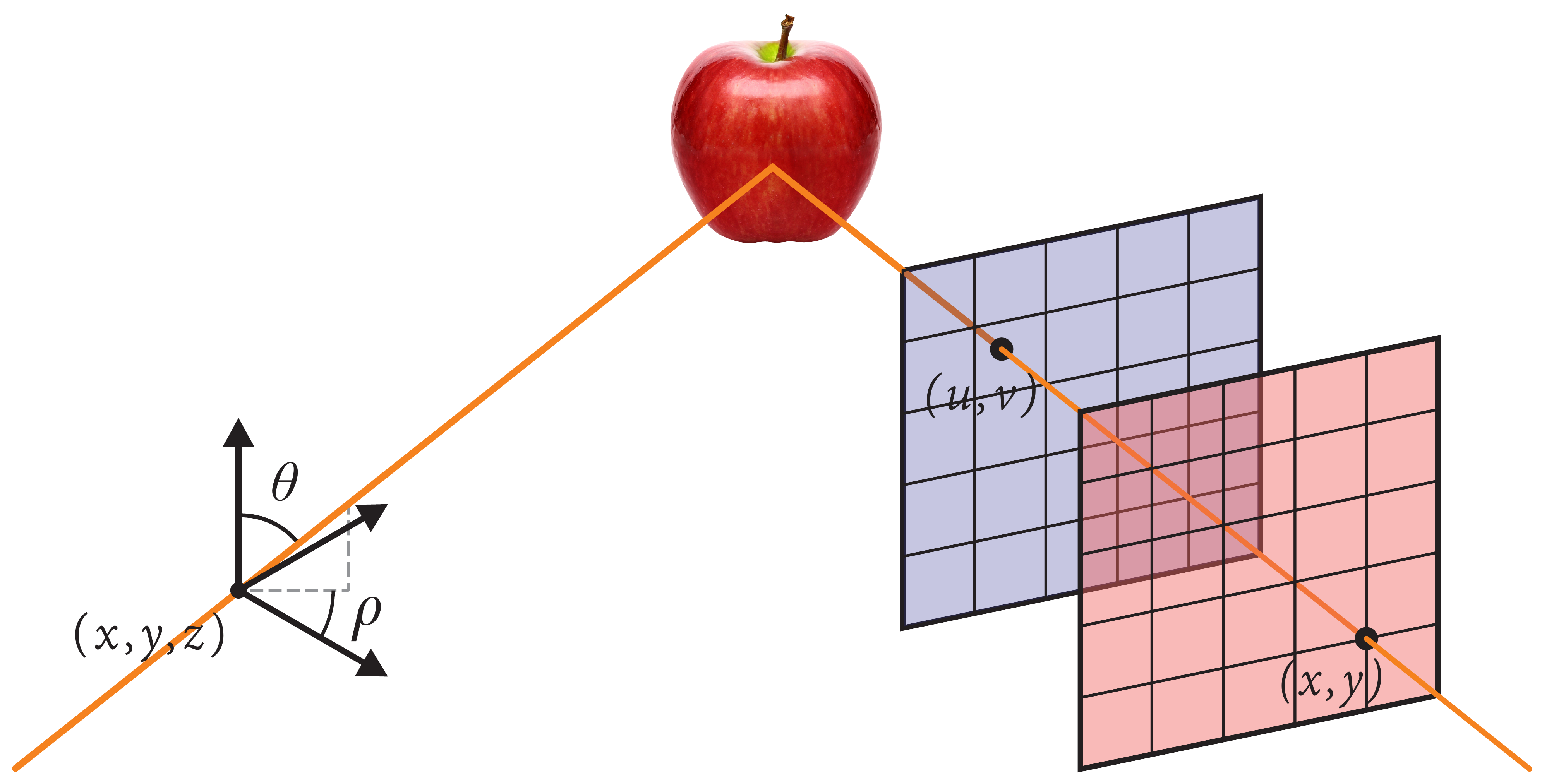}}
	\caption{5D Plenoptic function on the left versus two-plane parametrization on the right (4D \ac{lf} function).}
	\label{fig:2parallel-planes}
	\vspace{-5mm}
\end{figure}

\IEEEPARstart{T}he idea of the light flows through environment interpreted as a field was first established by Michael Faraday in 1846. The mathematical formalisation was proposed 28 years later by James Clerk Maxwell with his famous equations. The concept of \ac{lf} was then first defined in Arun Gershun's paper \cite{ArunTheLF} as the amount of light traveling in every direction through every point in 3D space. This amount of light is radiance, denoted by $L$, is measured in watts per steradian per meter squared. The {\it plenoptic function}  gives the radiance along all such arrays in a scene of 3D space with constant illumination
\begin{equation}
P(x, y, z, \theta, \phi, \lambda, t),
\end{equation}
the rays in space are parameterized by 3D coordinates $(x, y, z)$, two angles ($\theta, \phi$), wavelength $\lambda$ and time $t$. This 7 dimensional (7D) {\it plenoptic function} can be simplified into a 5D function where the time is sampled to the device frame rate and the wavelength is composed of 3 Red-Green-Blue (RGB) components. Assuming that the air around the object does not reflect or absorb the light and all ray intensities remain constant along their path, each ray is described by its intersection with two parallel planes denoted in this paper by $(u, v)$ and $(x, y)$ as illustrated in Fig.~\ref{fig:2parallel-planes} for angular and spacial coordinates, respectively. This 4D \cite{Levoy_1996_Graph} Light Field function $L(x, y, u, v)$ can be represented as a collection of perspective images of the $(x, y)$ plane viewed from a position on the $(u, v)$ plane.    

The \ac{lf} acquisition is performed by sampling both spatial and angular dimensions. The acquisition devices fall into two main categories depending on whether {\it camera arrays} or {\it plenoptic camera} acquisition technology is used. The camera arrays are matrices of synchronized cameras arranged in a plane often at regular interval, where each camera represents an angular sample and each image to spatial samples. Plenoptic camera relies on microlenses to capture lights coming from different directions. The spatial resolution is determined by the number of microlenses, while the angular resolution depends on the number of pixels behind each microlens. The resulting \ac{lf} image from plenoptic camera is then a collection of microlens images. This latter representation, called \ac{mi}, can be de-multiplexed in order to obtain \acp{sai}, where each \ac{sai} gathers pixels with the same relative position in the microlens image. The baseline of the \ac{lf} image captured by plenoptic camera is smaller\footnote{Plenoptic camera is also called narrow baseline plenoptic camera.} compared to the one captured by camera arrays.  
\begin{figure*}[t!]
	\centerline{\includegraphics[width=1\textwidth]{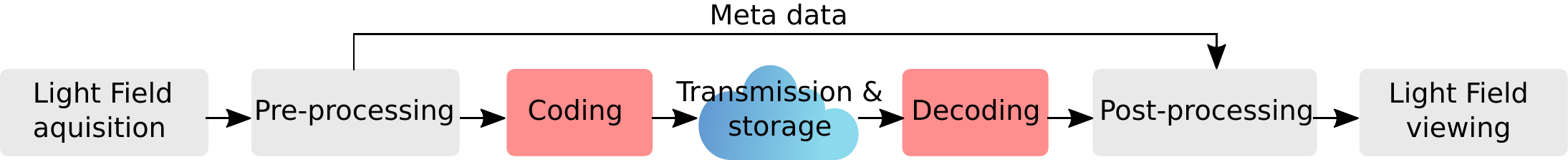}}
	\caption{Processing chain of light field technology from acquisition to end-user viewing.}
	\label{fig:LF-process}
\end{figure*}                      

The \ac{lf} image records important information about the scene geometry that can be leveraged in many applications. It enables for instance to simulate a change of a viewpoint for static or dynamic observer which can also enhance the viewing experience in \ac{vr} applications~\cite{10.1145/2766922}. The dense \ac{lf} can also enable high-quality depth map estimation~\cite{LF-depth-IP, 7298762} that can be used in the construction of an accurate point clouds~\cite{8683548,10114524619122461926} and image rendering with varying depth of field and focus plane post-acquisition~\cite{6831809}.

Fig.~\ref{fig:LF-process} illustrates the processing chain for \ac{lf} image deployment. After acquisition, the \ac{lf} image is processed by a pre-processing block to rearrange the data in an appropriate format for coding. The coding block removes spatial and angular redundancies in the \ac{lf} image to reduce the data size for efficient storage and transmission. The decoding block recovers from the bitstream the \ac{lf} image which is then processed by the post-processing block. This latter may  perform calibration, color correction with associated meta data or creating new interpolated views, synthetic aperture, refocusing, and extended focus for visualisation by the viewing block. The \ac{lf} image creates a large amount of data raising new challenges to the compression research community to design efficient coding solutions that drastically reduce the size of the \ac{lf} image while providing a high quality of experience in terms of immersion and realism offered by this technology. In response, several coding approaches have been proposed in the literature  which depend on the acquisition process of \ac{lf} image and its representation.   
\begin{figure}[h]
	\centerline{\includegraphics[width=0.49\textwidth]{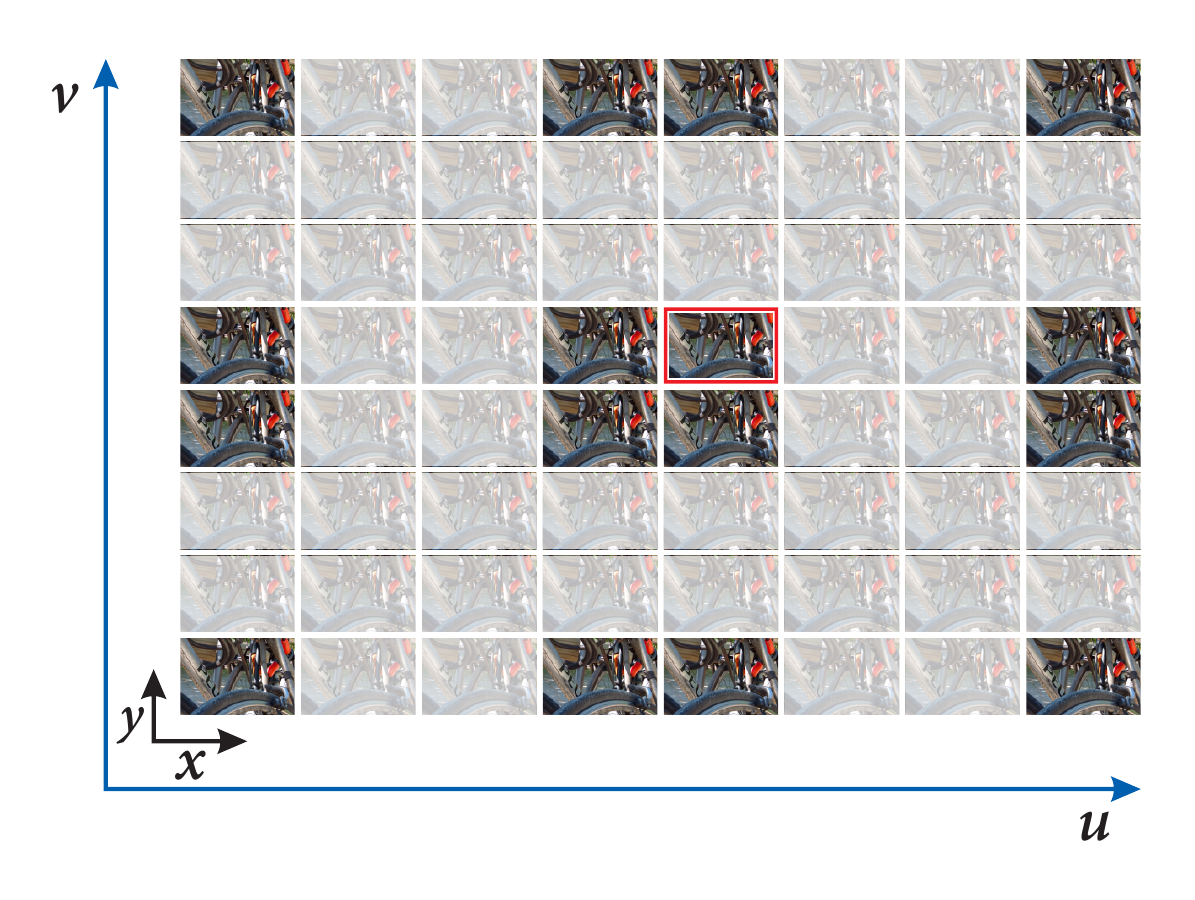}}
	\caption{Sparse representation of the $8  \times 8$ \ac{lf} image in subaperture representation with 16 reference views including the center view highlighted in red. }
	\label{fig:sparse-rep}
\end{figure}

\begin{figure}[t!]
	\centerline{\includegraphics[width=0.45\textwidth]{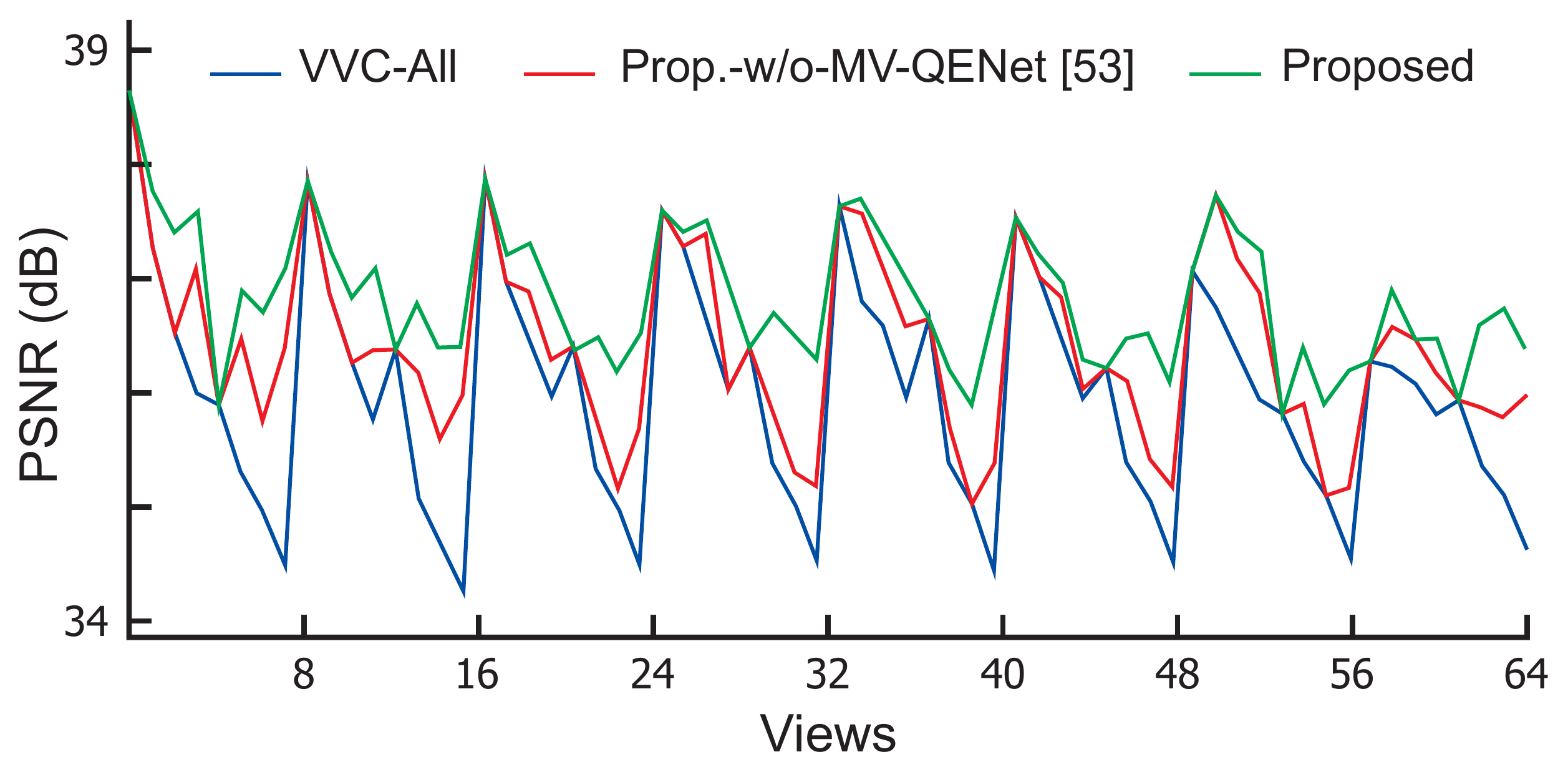}}
	\caption{Illustration of the quality fluctuation of the reconstructed \ac{lf} views before and after performing the quality enhancement block ({\it Stone-Pillars-Outside} test \ac{lf} image encoded at 0.0075 bpp).}
	\label{fig:LF-quality-fluc}
	\vspace{-2mm}
\end{figure}

In this paper, we investigate a lossy coding of lenslet-based acquisition \ac{lf} image, also known as \ac{llf} imaging. In the \ac{lf} processing chain illustrated in Fig.~\ref{fig:LF-process}, our contributions build the coding and decoding blocks that process the input \ac{llf} image and the encoded bitstream, respectively. The \acp{sai} (referred here to as views) of the \ac{lf} image are first arranged in a pseudo-video sequence, which is then encoded with the latest \ac{vceg}/\ac{mpeg} video coding standard called \ac{vvc} in temporal hierarchical coding configuration (i.e., temporal scalability). Only a sparse set of reference views, illustrated in Fig.~\ref{fig:sparse-rep}, are encoded at low temporal layers and are then used as reference to whether encode or synthesize the rest of views at the decoder side. We propose a \ac{lf-d2gan} to synthesize the missing views at the decoder side. The \ac{lf-d2gan} consists of one generator and two discriminators. The generator is composed of two components to predict the disparity and colors of a missing \ac{lf} view.  To enhance the generator performance, the training process is guided by two discriminators combining \ac{kl} and reverse \ac{kl} divergences into a unified objective function. Furthermore, in order to avoid large fluctuations quality across the reconstructed views, we propose a \ac{mv-qenn} as post-processing to propagate the quality from views decoded at high quality to other views. Fig.~\ref{fig:LF-quality-fluc} illustrates in red and green curves the quality fluctuations in \ac{psnr} of the \ac{lf} views reconstructed before and after performing the quality enhancement block, respectively. 
The performance of the proposed solution has been extensively assessed and compared with the state-of-the-art solutions. The experimental results showed the superiority of the proposed approach in terms of both coding efficiency and visual quality. 

The rest of this paper is organized as follows. Section~\ref{sec:related} gives a  review on existing \ac{lf} image coding solutions. The proposed solution is then described in Section~\ref{sec:proposal}. The performance of the proposed solution is assessed and analyzed in Section~\ref{sec:experimental} in terms of both coding efficiency and complexity. Finally, Section~\ref{sec:colusion} concludes the paper.

\begin{table*}[t!]
\renewcommand{\arraystretch}{1.1} 
	\begin{center}
		\caption{Main features of the existing \ac{llf} image coding solutions. 
		}
		  \resizebox{\linewidth}{!}{
		\label{Tab:LFcodingReview}
		{
			\renewcommand{\baselinestretch}{1}\footnotesize
			\begin{tabular}{|l|c|c|c|c|l|c|}
				\cline{1-7}
				{Solution} & {Fidelity} & {Representation} & {Geometry} & Sparsity & {Coding approach} & Standard-compl.  \\
				\cline{1-7}
				Viola {\it et al.}~\cite{2018_Graph_LFC} & Lossy  & \acp{sai} & \cmark & \cmark & \acs{hevc} \& Graph-based representation and coding & \xmark  \\
                \rowcolor{Gray} De Carvalho {\it et al.}~\cite{8451684} & Lossy  & \acp{sai} & \xmark & \xmark &  4D-DCT transform \& Hexadeca-trees & JPEG Pleno \\
				Astola {\it et al.}~\cite{8611756} & Lossy & \acp{sai} & \cmark & \cmark & JPEG 2000 \& Wraping and Sparse prediction & JPEG Pleno \\  
				\rowcolor{Gray} R.A. Farrugia {\it et al.}~\cite{8802937} & Lossless  & \acp{sai}  & \xmark & \xmark & 4D wavelet transform & \xmark \\
				Ahmad  {\it et al.}~\cite{8297145} \& \cite{7972959} & Lossy  & \acp{sai} & \xmark & \xmark & Multi-View HEVC  & MV-HEVC \\
				\rowcolor{Gray} Conti {\it et al.}~\cite{CONTI2018144} & Lossy  & \acs{mi} & \xmark & \xmark & \acs{hevc} \& \ac{bi-ss} & \xmark \\
				{{Liu}} {\it et al.}~\cite{8793171} & Lossy  & \acs{mi} & \xmark & \xmark & \acs{hevc} \&  Gaussian Process Regression (GPR)-based prediction & \xmark \\
				\rowcolor{Gray} Jiang \textit{et al.}~\cite{8022889} & Lossy  & \acp{sai} & \cmark & \xmark & Homography-based Low Rank Approximation & \xmark \\
				 Dib {\it et al.}~\cite{8712652}  & Lossy  & \acp{sai} & \cmark & \xmark & Super-Ray Based Low Rank Approximation & \xmark\\
			    \rowcolor{Gray} Zhao {\it et al.}~\cite{7805595} & Lossy  & \acp{sai} & \xmark & \xmark & Pseudo-video sequence \& \acs{jem} codec & \xmark \\
				  Liu {\it et al.}~\cite{7574674} & Lossy  & \acp{sai} & \xmark & \xmark & Pseudo-video sequence \& \acs{jem} codec & \xmark\\
				\rowcolor{Gray} Hou {\it et al.}~\cite{Hou:2019:LFI:3347960.3347993}  & Lossy  & \acp{sai} & \xmark & \cmark & \acs{hevc} \& CNN-based angular super-resolution  & \xmark \\
				  Jia {\it et al.}~\cite{8574895} & Lossy & \acp{sai} & \xmark & \cmark & \acs{hevc} \& LF-GAN & \xmark \\
				 \rowcolor{Gray} Zhao {\it et al.}~\cite{8297146} & Lossy  & \acp{sai} & \cmark & \cmark & \acs{hevc} \& Linear Approximation & \xmark \\
				    Bakir {\it et al.}~\cite{8712614} & Lossy  & \acp{sai} & \cmark &  \cmark & \acs{hevc}, Linear Approximation and CNN   & \xmark \\
				  \rowcolor{Gray} Wang {\it et al.}~\cite{9105769} & Lossy  & \acp{sai} & \xmark & \cmark & \acs{hevc} \& Multibranch Spatial Transformer Networks & \xmark \\
				    Komatsu {\it et al.}~\cite{8451812}  & Lossy  & \acp{sai} & \cmark & \cmark & Binary images representation of the \ac{lf} & \xmark\\
           	\rowcolor{Gray} {{Chen}} {\it et al.}~\cite{9120185} & Lossy  & \acp{sai} & \xmark & \cmark & \acs{hevc} \&  Global Multiplane Representation  & \acs{hevc} \\
		  {Conti} {\it et al.}~\cite{8338049} & Lossy  & \acp{sai} & \xmark & \cmark & \acs{hevc} \&  field-of-view scalability & \xmark \\
			\rowcolor{Gray}	{{Zhao}} {\it et al.}~\cite{8769976} & Lossy  & \acp{sai} & \xmark & \cmark & MV-\acs{hevc} \&  Super Resolution \acs{cnn} & MV-\acs{hevc} \\
		 	Proposed & Lossy  & \acp{sai} & \xmark & \cmark & \ac{vvc}, \ac{lf-d2gan} and \ac{mv-qenn}  & \ac{vvc} \\
				\hline
		\end{tabular}}}
	\end{center} 
	\vspace{-6mm}
\end{table*}

\section{Related Work}
\label{sec:related}
To ensure efficient storage and transmission of \ac{llf} imaging, many coding solutions have been proposed in recent years. In this section, we will give a brief review on \ac{llf} image coding solutions available in the literature. For more exhaustive description of these solutions, the reader may also refer to two overview papers recently published in~\cite{9016088} and ~\cite{9020136}. Table~\ref{Tab:LFcodingReview} summarises the main features of the covered solutions in terms of fidelity (lossy or lossless coding), data representation prior encoding, consideration of the geometry and sparse representation, the adopted coding approach, and finally the compliance with a coding standard. The geometry-related data can represent the distances of a 3D scene such as depth or disparity information, when not available at the decoder, this geometry-related data can be estimated from the decoded views (texture).               

Authors in~\cite{CONTI2018144} proposed a \ac{bi-ss} estimation and compensation to remove spacial and angular redundancies within the \ac{lf} image in \ac{mi} representation. The \ac{bi-ss} prediction is proposed as an additional prediction mode under the \ac{hevc} encoder in Intra coding configuration. The \ac{bi-ss} prediction performs whether uni-predictive or bi-predictive coding with reference blocks from already decoded and filtered (in-loop) blocks in the same image. The bi-predictive coding performs a weighed combination of two candidate blocks as in \ac{hevc} bi-directional prediction but the reference blocks are from the same image. The best coding mode among the 35 Intra prediction modes and the two new modes is selected by the encoder through a regular rate-distortion optimization process. 

A two streams coding scheme has been proposed by Viola {\it et al.} in
~\cite{2018_Graph_LFC} for \ac{lf} image in subaperture representation. The \ac{lf} views are first split in two sets (streams). The views of the first set (reference views) are arranged in a pseudo-video sequence which is then encoded with \ac{hevc} encoder in Inter configuration (low delay configuration). The encoder also estimates the graph that models the dependencies among the \ac{lf} image views. The graph weights are quantized and transmitted to the decoder. This latter decodes the reference views and with the decoded graph, it solves an optimization problem to recover the rest of views, and at low bitrate, enhance the quality of the reference views.  

A 4D separable transform through the 4D-\ac{dct} is used in~\cite{8451684} to decorrelate the \ac{lf} image and concentrate its energy in few coefficients. These coefficients are then clustered using hexadeca-tree structure, where each node corresponds to a 4D block of transform coefficients in a specific sub-band. Each node in this tree structure can be further sub-divided into sixteen children (sub-regions) and a binary symbol \textit{1} is encoded otherwise \textit{0} is encoded for no further split. The decision to terminate the recursive split process is taken when a sub-region contains only zero coefficients or a single non-zero coefficient. The binary symbols of the constructed hexadeca-tree with the quantized \ac{dc} and \ac{ac} \ac{dct} coefficients are encoded by a \ac{cabac} with three contexts. One binary context is used for segmentation flags and two non-binary contexts for \ac{dc} and \ac{ac} coefficients in each sub-band. This solution has been farther enhanced in terms of both coding efficiency and random access feature. The solution was adopted by the JPEG Pleno standard as the 4D transform mode.  

Authors in~\cite{8611756} proposed a $N$-layer hierarchical coding scheme for \ac{lf} image in subaperture representation. The \ac{lf} views are first arranged in $N$-layer structure, where $N$ is set to 6. The $N-1$ first layers are called reference layers since the associated views are used as reference to encode views at higher layers. The views of the first layer with the corresponding inverse depth maps\footnote{The inverse depth map corresponds to the ratio between the camera focal and the pixel depth value.} are encoded with \hbox{JPEG 2000}~\cite{10.5555/2588198}. The inverse depth map of a view at higher layer is synthesized from the reference inverse depth map with a simple pixel-wise wrapping operation. The reference view candidate is then wrapped to the location of the view to generate the so-called wrapped reference view candidate. The wrapped reference views are fused in a single high quality reference view. The fusion is performed by a simple least-squares regression technique. The coefficients derived from this latter step are quantized in 16 bits and sent to the decoder. A least-squares minimization method is also used to predict the encoded view from the constructed reference view and the resulting non-zero coefficients for each color component are encoded with an arithmetic encoder. Finally, the prediction residue, which is the difference between the view to encode and the merged reference view, is encoded with JPEG 2000. This solution has also been adopted by JPEG Pleno standard as the 4D predictive mode.             
 
Ahmad {\it et al.}~\cite{8297145} proposed to arrange the \ac{lf} image as a multi-view sequence that is encoded by the Multi-View extension of \ac{hevc} standard (MV-\ac{hevc})~\cite{7351182}. A row of \acp{sai} as shown in Fig.~\ref{fig:sparse-rep} corresponds to a single view in the multi-view sequence. Temporal and multi-view predictions are used to efficiently leverage the spatial and angular correlations in the \ac{lf} image. Four hierarchical predictions levels with specific \acp{qp} were defined in horizontal (i.e., temporal) and vertical (i.e., views) directions to perform efficient prediction of the \acp{sai}. This solution has been described in more details and assessed under the  JPEG Pleno test conditions in~\cite{8853251}. 
The MV-\ac{hevc} extension has also been used in~\cite{7972959} to encore the \acp{sai} arranged in four quadrants. All views are encoded one quadrant after another to reduce the reference buffer size. Under each quadrant, a hierarchical coding configuration is used to leverage angular and spatial correlations within the views.   
 
Jiang \textit{et al.}~\cite{8022889} proposed a coding method called homography-based low rank approximation (HLRA). This method jointly optimizes global or multiple homographies that align the \ac{lf} views and low rank approximation matrices. Global or multiple homographies configuration is selected depending on the variation of the disparity across the views. The low-rank representation of the \ac{lf} image is then encoded with  \ac{hevc}. Dib \textit{et al.}~\cite{8712652} proposed a compression scheme for \ac{lf} image using super-ray based local low rank models. A novel method for disparity estimation and compensation was proposed so that the super-rays are constructed to yield the lowest approximation error for a given rank. This representation is based on two low rank models, one for the central view pixels that are visible in all views while the other is used for occlusions. Authors in ~\cite{8451812} proposed a new coding concept for 4D \ac{lf} relying on a new representation of the \ac{lf} with $N$ binary images and the corresponding weights.  A set of binary basis images is selected to capture a common structure among all viewpoints, and the difference among the viewpoints are represented with pixel-independent weight. A least squares problem is solved to derive the $N$ binary images and the corresponding weights. These images can then be encoded with an arithmetic encoder. 

Several works~\cite{7805595},\cite{7574674} have investigated a straight forward coding approach that organizes the \ac{lf} views in a pseudo-video sequence, which is then encoded with a classical hybrid video encoder. For instance, Liu \textit{et al.}~\cite{7574674} proposed a compression of \ac{lf} image based on pseud-video sequence of \acp{sai}. A subset of views is then arranged in a specific coding order that accounts for similarities between adjacent views and encoded using the \ac{jem} encoder.

Another approach consists in encoding a spare set of views using a video encoder, while the rest of views are synthesized at the decoder side. The latter solution has been followed by several authors~\cite{8297146,8712614,8574895, Hou:2019:LFI:3347960.3347993, 9105769, 8451812, 9120185, 8338049, 8769976}, for instance, linear approximation has been investigated in~\cite{8297146} to estimate the views at the decoder from neighbour views, while a combination of linear approximation and \ac{cnn} has been proposed in~\cite{8712614} to synthesize missing views at the decoder side. In the same way, Jia {\it et al.}~\cite{8574895} proposed to use the \ac{gan} to generate unsampled views. To enhance the coding efficiency, the authors proposed to encode and transmit the residual error between the generated uncoded views and their original versions.  Hou {\it et al.}~\cite{Hou:2019:LFI:3347960.3347993} proposed a method that exploits the inter- and intra-view correlations effectively by characterizing its particular geometrical structure using both learning and advanced video coding techniques. The \acp{sai} are first partitioned into key and non-key \acp{sai}. The key \acp{sai} are encoded with a 2D video encoder while the non-key images are synthesized at the decoder side by a learning-based angular super-resolution approach. The residual images between the original non-key \ac{sai} and their synthesized versions are also arranged in a pseudo-video sequence and encoded with a  video encoder. Wang \textit{et al.}~\cite{9105769} proposed a novel light field image compression scheme using \ac{mstn} based view synthesis. First, a sparse subset of views, arranged into a pseudo-video sequence, are encoded by a video encoder. Then, the rest of views are synthesized based on the similarity between neighboring views with the \ac{mstn} block. This latter enables better characterization of the non-linear relationship between the sub-views with adaptive learning of the affine transformations between the neighboring views, which are used to warp the input views to generate accurate high-order approximation of the missing views. In~\cite{9120185}, the authors proposed to encode a set of reference views with \ac{hevc} while the rest of views are estimated at the decoder site in two steps.  The first step predicts a disparity-based global representation and then the prediction is performed as a second step based on multiplane as the form of this global representation. 
The reference views are encoded in \cite{8769976} with the MV-\ac{hevc} standard. The quality of these views is first enhanced at the decoder side with a quality enhancement \ac{cnn}. The resulting enhanced views are then used to as input to predict the reset of views with two super-resolution \acp{cnn}.

\begin{figure*}[t]
	\centerline{\includegraphics[width=\textwidth]{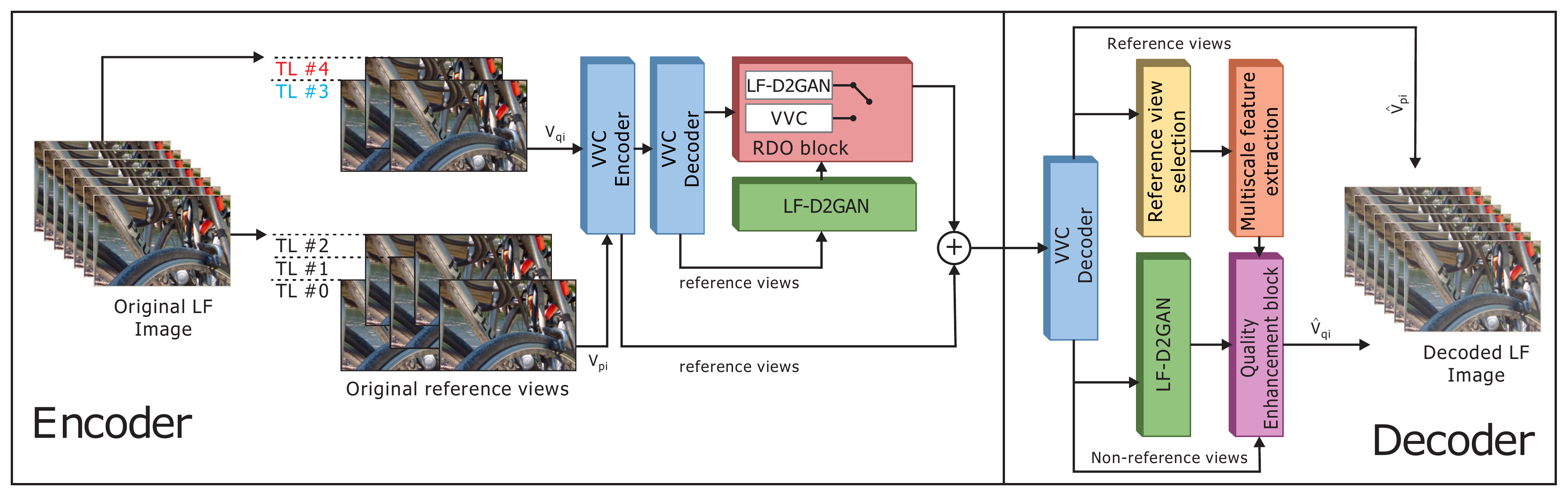}}
	\caption{Overall scheme of the proposed coding solution.}
	\label{fig:Schema}
\end{figure*}

All  these  latest \ac{lf}  coding  solutions based on view synthesis exploit  a  sparse representation  of  the  \ac{lf}  image  and  differ in  the  process of selecting the key of views and especially in the view synthesis algorithm that can be~\cite{9020136}: 1) an Depth Image Based Rendering (DIBR) based method, 2) a transform-assisted method and 3) a learning-based method. The \ac{lf} coding solutions using learning-based view synthesis obtained the highest coding performance compared to other view synthesis techniques. However, these solutions suffer from three major drawbacks: first, the used synthesis module is based on a learning approach generally relying on a variant of \ac{cnn} to synthesize the discarded views. Hence, a learning-based approach shows  a  large variation in performance for a set of \ac{lf} images with different color, spatial and occlusion characteristics. Second, in most of the proposed methods, the number of encoded/dropped views is predetermined and set manually, which does not reflect the best selection choice that leads to the highest coding efficiency. Finally, we  noticed a large visual quality fluctuation across the reconstructed views at the decoder side. This quality fluctuation may result in lower viewing experience in many \ac{lf} applications. 

Our proposed solution belongs to the learning-based view synthesis method class and has been designed to overcome the mentioned drawbacks with the following main contributions:    
\begin{itemize}

\item A use of \ac{vvc}'s temporal scalability structure  to drop views without impacting the decoding of other ones and without signaling them in the bitstream, thus keeping the bitstream compliant with the \ac{vvc} codec.
\item  A learning-based view synthesis method called \ac{lf-d2gan} which is based on two \ac{cnn}s for color and disparity estimations that are simultaneously trained with two adversarial discriminators. 
\item A rate-distortion optimization algorithm that selects at the encoder side whether the non-reference views should be encoded with \ac{vvc} or synthesized at the decoder side by the \ac{lf-d2gan} block. \item A novel \ac{mv-qenn} method which is applied as a post-processing to improve the quality of the non-reference views. This \ac{mv-qenn} block propagates the quality to the target non-reference view from two carefully selected reference views. The \ac{mv-qenn} block allows reducing the quality fluctuation across the decoded \ac{lf} views and thus increases the viewing experience.
\end{itemize}

\section{Proposed Method}
\label{sec:proposal}
In this paper, we propose a novel approach to encode a \ac{lf} image in subaperture representation. The $N$ \acp{sai} (views) are first split into a sparse set of $N_R$ reference views ($V_{p_1}, \dots, V_{p_{N_R}}$) and $N-N_R$ non-reference views ($V_{q_1}, \dots, V_{q_{N-N_R}}$), with $p_i$ and $q_j$ are the angular $(u,v)$ positions of the reference and non-reference views, respectively. All these views are arranged in a pseudo-video sequence which is then encoded with a hybrid Intra/Inter video encoder in hierarchical coding configuration (i.e., temporal scalability). The reference views are encoded at low temporal layers and are used as reference for encoding the non-reference views. These latter are encoded at higher temporal layers and thus are not used as reference to encode the reference views. The non-reference views are also synthesized at the encoder side with a synthesis block that takes as input the decoded reference views. The encoder then performs a \acl{rdo} between the synthesized and decoded non-reference views and selects the one that minimizes the rate-distortion cost. The bitstream is therefore composed of reference views and a set of non-reference views encoded with a video encoder. The non-reference views for which the rate-distortion cost is lower with the synthesis block are discarded from the bitstream  without impacting the decoding of the transmitted views. The decoder performs inverse encoding operation to decode the transmitted views. The non-reference views dropped by the encoder are then synthesized and, thereafter, a quality enhancement is performed on them as a post-processing to  ensure consistency of quality between views.

The block diagram of the proposed \ac{lf} image coding scheme is illustrated by Fig.~\ref{fig:Schema}. In the rest of this section we will investigate in more details the elementary blocks of our proposed approach including the 2D video encoder, view synthesis, rate-distortion optimization and post-processing quality enhancement.

\subsection{LF pseudo-video sequence encoding}
\label{subsec:VVC}
The \ac{lf} image presents large angular and spatial correlations in the SAIs. These SAIs when arranged in a pseudo-video sequence can be efficiently encoded with a hybrid video encoder that leverages these correlations through Intra/Inter predictions and transform coding. The \ac{jvet}, jointly established by ISO/MPEG and ITU/VCEG standardisation committees, has released in July 2020 the latest video coding standard called \ac{vvc}~\cite{Wien2017}. \Ac{vvc} enables a bitrate saving of 35\% to 50\% with respect to its predecessor \ac{hevc} standard for the same visual quality~\cite{PCSVVC}. This coding gain is enabled by several coding tools at different levels of the coding chain including frame partitioning, Intra/Inter predictions, transform, quantization, entropy coding and in-loop filters. In particular, \ac{vvc} performs more efficient Intra and Inter predictions than \ac{hevc} by either enhancing \ac{hevc} tools or introducing new ones~\cite{jvetdoc,8281012,7906344}.

The proposed approach is based on the \ac{vvc} standard to encode the pseudo-video sequence in temporal scalabilty configuration. However, our approach is codec agnostic in that it can be  used with any 2D video codec that supports hierarchical \ac{gop} structure and temporal scalability.

\begin{figure}[t]
	\centerline{\includegraphics[width=0.49\textwidth]{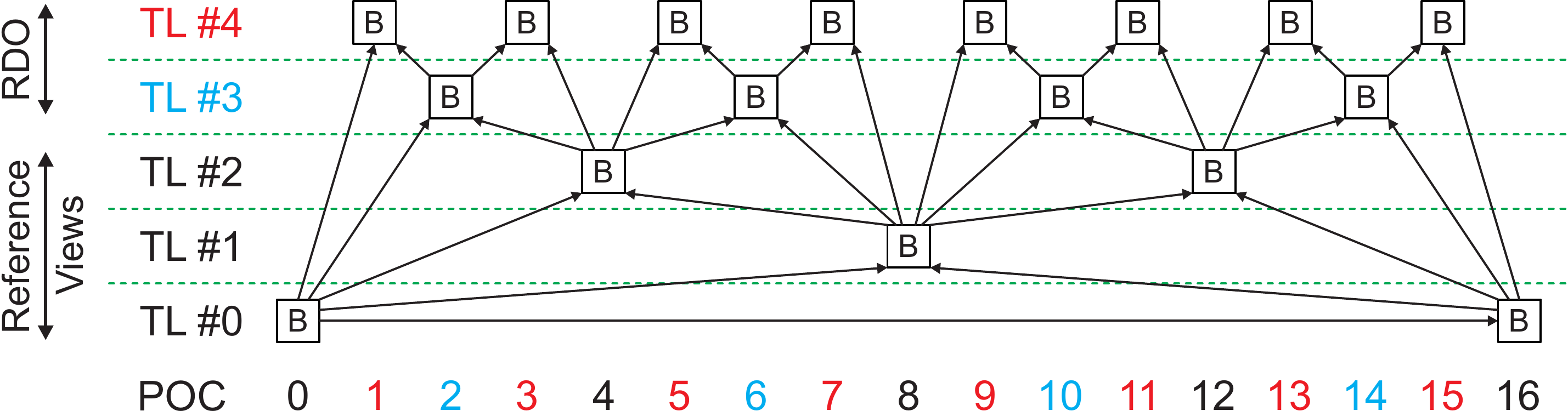}}
	\caption{Hierarchical prediction structure in \acs{vvc} in \acs{ra} coding configuration.}
	\label{fig:VVCGOP}
	\vspace{-3.5mm}
\end{figure}

The advanced Intra/Inter \ac{vvc} tools will take advantage of the  spatial and angular redundancies of the \ac{lf} image. \Ac{vvc} supports by design temporal scalability through the \acl{ra} coding configuration. This latter, illustrated in  Fig.~\ref{fig:VVCGOP}, enables different temporal layers and each temporal layer uses as reference only frames from lower temporal resolution, i.e., lower layer. Therefore, frames of each temporal layer $t_i$ can be dropped without impacting the decoding of frames at lower temporal resolution $t_j$ with $t_i>t_j$. In the proposed coding approach, we leverage the concept of temporal resolution to drop views at the encoder without impacting the decoding process and thus performing the best rate-distortion performance.

\subsection{LF Dual Discriminator Generative Adversarial Nets}
\label{subsec:D2GAN}
Several solutions \cite{Silhouette-2011, ChaurasiaDepthSynthesis2013, Kalantari:2016:LVS:2980179.2980251} have been proposed in the literature to synthesize a novel or missing \ac{lf} view. The problem in the coding scheme consists in estimating a missing view $\tilde{V}_q$ from a sparse set of decoded $N_R$ reference views $\tilde{V}_{p_1}, \tilde{V}_{p_2}, \dots, \tilde{V}_{p_{N_R}}$ 
\begin{equation}
\tilde{V}_q = f \left (\tilde{V}_{p_1}, \tilde{V}_{p_2}, \dots, \tilde{V}_{p_{N_R}}, q \right ),
\end{equation}
where $p_1, p_2 \cdots, {p_{N_R}}$ and $q$ are the $(u, v)$ positions of the $N_R$ reference views and the estimated non-reference missing view $\tilde{V}_q$, respectively. 

\begin{figure*}[t]
	\centerline{\includegraphics[width=1\textwidth]{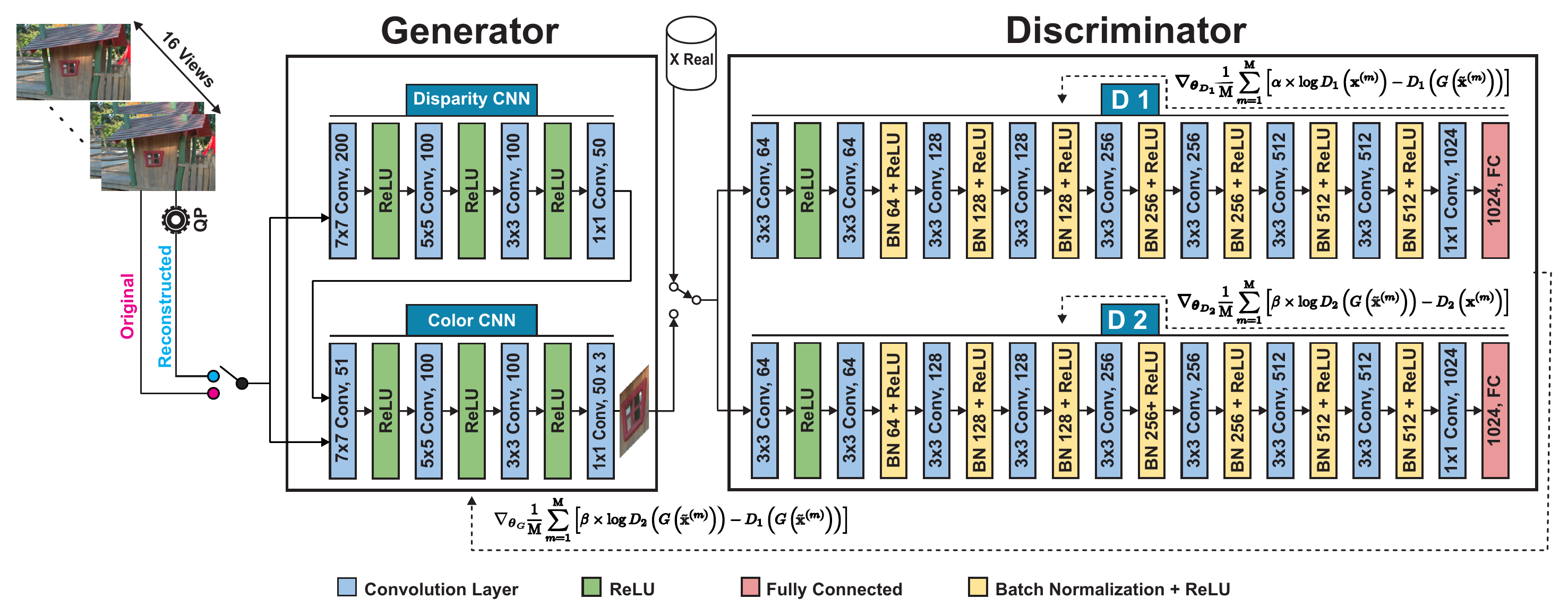}}
	\caption{Architecture of the proposed \acf{lf-d2gan}.}
	\label{fig:D2GANArchitecture}
\end{figure*}

Inspired by the success of CNN architectures for \ac{lf} view synthesis~\cite{Kalantari:2016:LVS:2980179.2980251}, we propose to use a learning-based approach to synthesize the missing views at the decoder side. The proposed synthesis block is composed of one generator $G$ and two discriminators $D_1$ and $D_2$. Similar to~\cite{Kalantari:2016:LVS:2980179.2980251}, the generator is broken-down into two \acp{cnn} for efficient estimation of disparity and color, as illustrated in Fig.~\ref{fig:D2GANArchitecture}. These two sequential components are trained simultaneously to minimize a cost function. Our contribution in this block consists in enhancing the performance of the generator by conducting unsupervised learning guided by two discriminators. The disparity CNN estimates the disparity of the missing view $D_q$ from a set of features $K$ computed from the input reference views 
\begin{equation}
D_q = g_d \left (K \right )
\end{equation}
where $g_d$ is the function that computes the relationship between the input features and the disparity of the target view. The input features $K$ consist mainly of mean and standard deviation of input reference views wrapped at different disparity levels.    

Using the estimated disparity $D_q$, the reference views are then wrapped to the target view 
\begin{equation}
    \bar{V}_{p_i}(s) = \tilde{V}_{p_i} \left [ s + (p_i - q ) D_q (s) \right ],
\end{equation}
where $s$ is the $(x, y)$ pixel position. 

The $N_R$ wrapped reference views $\bar{V}_{p_1}, \bar{V}_{p_2}, \dots, \bar{V}_{p_{N_R}}$ are provided to the color estimation \ac{cnn} $g_c$ in order to estimate the color of the missing view. The color \ac{cnn} estimates the missing view by using all wrapped reference views $\bar{V}_{p_1}, \bar{V}_{p_2}, \dots, \bar{V}_{p_{N_R}}$, its disparity map $D_q$ estimated by the disparity CNN and its position $q$.
\begin{equation}
    \tilde{V}_{q} = g_c \left (\bar{{V}}_{p_1}, \dots, \bar{{V}}_{p_{N_R}} , D_q, q \right ).
\end{equation}

As mentioned in Section~\ref{sec:intro}, the proposed coding approach is based on the \ac{lf-d2gan} block. \Acp{gan} are deep neural net architectures composed of two consecutive neural network models, namely generator $G$ and discriminator $D$. \Ac{gan} enables to simultaneously train the two models: the generative model $G$ that captures the data distribution, and the discriminative model $D$ that estimates the probability that a sample came from the training data rather than from the generator $G$~\cite{goodfellow2014generative}.  \Ac{gan} has recently achieved great success in various fields, especially in fake video generation, \ac{lf} super-resolution and objects detection~\cite{{8099502},{Bai_2018_ECCV}}. 

The training of the two generators $g_d$ and $g_c$ is guided by two discriminators $D_1$ and $D_2$. Given an input data $x$ which consists here in an input data patch, the first discriminator $D_1$ rewards a high score for real data ($\mathbb{P}_{train}$) and returns low score for data generated by the generator ($\mathbb{P}_{G}$). In contrast, the second discriminator $D_2$ returns low score when the input data follows the real data distribution and high score for the input data close to the model distribution. The two generators are then trained simultaneously to generate samples that fool the two discriminators in a three-player minimax optimization game
\begin{equation}
\label{eq:D2GAN}
\begin{split}
&\min_{\theta_G} \max_{\theta_{D1},\theta_{D2}} \mathcal{L}~(\theta_G,\theta_{D1},\theta_{D2}) = \alpha \, \mathbb{E}_{x \backsim \mathbb{P}_{train}} [\log{D_{1}( x)}] \\
& + \mathbb{E}_{\tilde{x} \backsim \mathbb{P}_{G}} [-D_{1}(G(\tilde{x}))] + \mathbb{E}_{x \backsim \mathbb{P}_{train}} [-D_{2}(x)] \\
& + \beta \, \mathbb{E}_{\tilde{x} \backsim \mathbb{P}_{G}} [\log{D_{2}(G(\tilde{x}))}], 
\end{split}
\end{equation} 
where $\mathbb{E}$ represents expected value, $x$ is the real data, $\tilde{x}$ is the generated data, $\mathbb{P}$ represents the probability distribution, $\alpha$~and~$\beta$ are two hyper-parameters (0  $< \alpha$, $\beta \leq$ 1) to stabilize the learning of the model and control the effect of \ac{kl} and reverse \ac{kl} divergences on the optimization problem~\cite{NIPS2017_6860}. The models are trained by alternatively updating discriminators parameters $\theta_{D_1}$, $\theta_{D_2}$ and the generator parameters $\theta_D$ by solving a minimax optimization game.

Three cost functions defined in \eqref{eq:CostD1}, \eqref{eq:CostD2} and \eqref{eq:CostG} are computed to obtain the error that should be transmitted respectively to $D_{1}$, $D_{2}$ and $G$ for their backward weights updating, as shown in  Fig.~\ref{fig:D2GANArchitecture} (dash lines). Thus, \eqref{eq:CostD1} and \eqref{eq:CostD2} are used to update the weights of the discriminators $D_{1}$ and $D_{2}$, respectively, by ascending the obtained stochastic gradient.
\begin{equation}
\label{eq:CostD1}
\begin{split}
\nabla_{\theta_{D1}} \dfrac{1}{M} \sum_{m=1}^{M}   \alpha \, \log{D_{1}(x^{(m)})} -D_{1}(G(\tilde{x}^{(m)})),
\end{split}
\end{equation} 
\begin{equation}
\label{eq:CostD2}
\begin{split}
\nabla_{\theta_{D2}} \dfrac{1}{M} \sum_{m=1}^{M}   \beta \, \log{D_{2}(G(\tilde{x}^{(m)}))} -D_{2}(x^{(m)}),
\end{split}
\end{equation}
while \eqref{eq:CostG} represents the cost function to gain the error that should be given to the generator $G$ for its weight updating.
\begin{equation}
\label{eq:CostG}
\begin{split}
\nabla_{\theta_{G}} \dfrac{1}{M} \sum_{m=1}^{M}   \beta \, \log{D_{2}(G(\tilde{x}^{(m)}))} -D_{1}(G(\tilde{x}^{(m)})).
\end{split}
\end{equation} 

\begin{figure*}[t]
	\centerline{\includegraphics[width=1\textwidth]{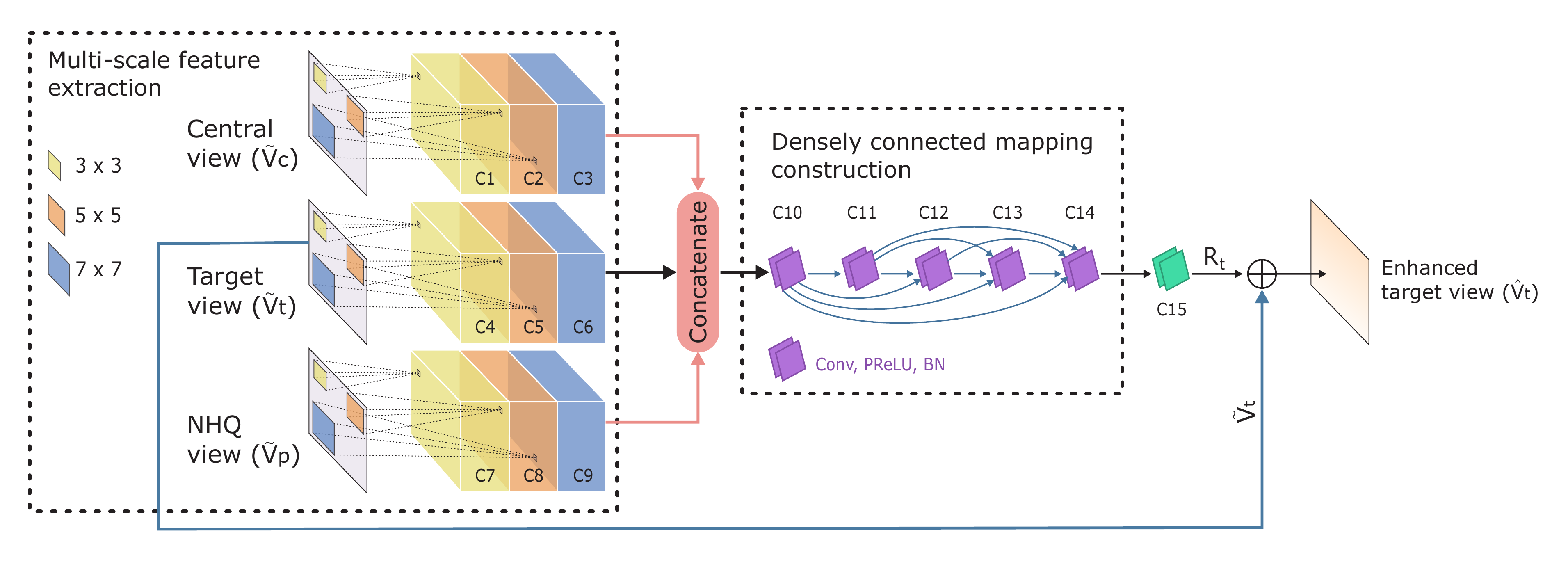}}
	\caption{Detailed architecture of the proposed \acf{mv-qenn}.}
	\label{fig:QualityEnhancementBlock}
\end{figure*}

\subsection{Rate-distortion optimization}
Instead of fixing  the number of dropped views, in our approach this is done adaptively on the basis of a \ac{rdo} process. At the encoder side, first, \ac{lf} subaperture views are organized into groups of 16 views that form \ac{gop}, as illustrated in  Fig.~\ref{fig:VVCGOP}. Next, in each \ac{gop}, the images of temporal levels 0, 1 and 2 are encoded using the \ac{vvc} codec, which constitute the reference views used in the synthesis process. Then, the images at the remaining levels 3 and 4 are either encoded using the \ac{vvc} codec or dropped. For these decisions, we propose a \ac{rdo} algorithm to select whether a non-reference view  should be encoded or synthesized at the decoder side. The proposed \ac{rdo} process is described in the Algorithm~\ref{Algo:rdo} and explained in the following.
\begin{algorithm}[t!]
	\caption{\acs{rdo} block based Lagrangian optimization}
	\label{Algo:rdo}
	\begin{algorithmic}
		\REQUIRE $\mathcal{J} \leftarrow \{\forall\ v \in TL\#[3\ or\ 4], \ \forall \ m \in \{VVC, LF-D2GAN\}, \mathcal{J} = D + \lambda \, R \}$
         
		\FORALL{ $v \in TL\#4$}
		
			\IF{$\mathcal{J}(VVC)\ <\ \mathcal{J}(LF\text{--}D2GAN)$}
		       \STATE Encode $v$ by VVC
		        \STATE Send($v$)

		     \ELSE
	        	\STATE Generate $v$ by LF-D2GAN
		    \ENDIF
		
		\ENDFOR
		\FORALL{ $v \in TL\#3$}
		
		\IF{$\mathcal{J}(VVC)\ > \ \mathcal{J}(LF\text{--}D2GAN)$ \AND all dependent views are synthesized by LF-D2GAN}
		\STATE Generate $v$ by LF-D2GAN
		\ELSE
		\STATE Encode $v$ by VVC
		\STATE Send($v$)
		\ENDIF		
		\ENDFOR		
	\end{algorithmic}
\end{algorithm}

As illustrated in Fig.~\ref{fig:VVCGOP}, we apply \ac{rdo} process on the 3 consecutive frames, i.e.,  frame $i$ at level 4, frame $i+1$ at level 3 and frame $i+2$ at level 4. It should be noted that if one of the views at temporal level 4 (frame $i$ or $i+2$) is encoded using \ac{vvc}, then the frame $i+1$ at level 3 must also be encoded using \ac{vvc}, because this layer will be used as a reference for the frames at temporal level 4. The main reasons behind only considering the 2 upper levels exclusively to the \ac{rdo} block are, firstly, after an extensive study, we found that these levels together represent around 28\% of the total bitrate. Second, the views at the upper levels are not used as references in the \ac{vvc} coding scheme to encode reference views. Thus, the proposed \ac{rdo} block can decide which views from the upper level can be encoded using \ac{vvc} or dropped and synthesized using \ac{lf-d2gan}. To reach this goal, the encoder computes the rate-distortion (RD) cost function $J$ given by \eqref{eq:cost_function} for both the \ac{vvc} decoded view and the one synthesized by the \ac{lf-d2gan}.
\begin{equation} \label{eq:cost_function}
\mathcal{J} = D + \lambda \, R,
\end{equation}
where $\lambda$ is the Lagrangian multiplier, $D$ is the distortion and $R$ is the rate in bits per pixel (bpp). To set the Lagrangian multiplier $\lambda$, we empirically determine its value by testing a large set of \ac{lf} images. We found that the value of $0.1$ for $\lambda$ is optimal and for which the Lagrangian optimization is giving the best performance. 

It should be noted that the dropped views are not signaled in the bitstream and the decoder can detect the missing views based on the \ac{poc} of the decoded frames in the \ac{gop}. The \ac{poc} of a non-reference view not present in the bitstream is identified and its angular position $q$ is sent to the \ac{lf-d2gan} block. This also has the advantage to make the bitstream compliant with the \ac{vvc} standard.

\subsection{Multi-View Quality Enhancement Net}
After analyzing the quality of each view at the output of the \ac{rdo} block, we noticed that there is a significant fluctuation in the quality of the decoded views. Specifically, we  have found that the non-reference views have lower quality than the reference ones. This quality fluctuation is caused by the high \ac{qp} values assigned to the views at high temporal layers in the \ac{vvc} hierarchical coding structure on the one hand, and the unpredictable output quality of the \ac{lf-d2gan} block on the other hand. Thus, we propose to perform a post processing on these non-reference views at the decoder side using \ac{mv-qenn} block  to further enhance their quality and reduce the quality fluctuation across \ac{lf} views, as shown in Fig.~\ref{fig:Schema}. Here, the concept of quality enhancement consists in predicting the residual errors $R_{q} = V_{q} - \tilde{V}_{q}$ of the non-reference views using a \ac{cnn}.  At the decoder side, the proposed \ac{cnn} architecture uses 3 views as an input. These three views include the target decoded non-reference view $\tilde{V}_t$, the decoded central view $\tilde{V}_c$ (which is of the highest quality as it is encoded in Intra using low \ac{qp} value) and one neighbor decoded reference view $\tilde{V}_{p}$. $\tilde{V}_{p}$ is selected among the reference views (except the central view already included in input) through a blind \ac{iqa} metric called  \ac{cornia}~\cite{Ye2012KmeansSR}. This latter has the advantage of providing image quality scores without access to reference images and showed a high correlation with humane appreciations. 

Thus, at the decoder, the \ac{rvs} block picks among 15 neighbors views the $\tilde{V}_p$ view with the highest quality score computed by \ac{cornia} metric. The \hbox{3 views} ($\tilde{V}_{t}$, $\tilde{V}_c$, $\tilde{V}_{p}$) are then fed to the \ac{mv-qenn} which extracts the multiscale characteristics of the views and constructs a densely connected mapping in order to predict the residual errors and transfer the quality of neighbor views to the target view
\begin{equation}
    \hat{V}_t = \tilde{V}_t + h_\phi (\tilde{V}_t, \tilde{V}_c, \tilde{V}_p),
\end{equation}
with $h_\phi$ is the parametric function of quality enhancement neural network and $\phi$ its trainable parameters.

The architecture of this neural network is composed of two key components: the multiscale feature extraction (denoted by layers C1-9 in Fig.~\ref{fig:QualityEnhancementBlock}) and the densely connected mapping construct (denoted by layers C10-14 in Fig.~\ref{fig:QualityEnhancementBlock}). Multi-scale features extraction takes as input two reference views ($\tilde{V}_c$, $\tilde{V}_{p}$) and one target non-reference view $\tilde{V}_{t}$. The spatial characteristics of these three views are extracted by multiscale convolutional filters. After feature extraction, all feature maps from the input three views are concatenated, then flow into the densely connected block component. After obtaining the feature maps of these three views, the densely connected architecture is applied to build the nonlinear mapping of feature maps in order to improve the residual part. In fact, there are 5 convolutional layers in the nonlinear mapping of the densely connected architecture. Each of them has 32 convolutional filters with size of 3$\times$3. In addition, dense connection~\cite{Huang_2017_CVPR} is adopted to encourage feature reuse, strengthen feature propagation and mitigate the vanishing-gradient problem. Moreover, a \ac{bn} is applied to all 5 layers after PReLU activation to reduce internal covariate shift, thus accelerating the training process.

\begin{table}[t!]
\renewcommand{\arraystretch}{1.2} 
	\begin{center}
		\caption{Convolutional layers of \ac{mv-qenn} block.}
		\label{tab:QualityEnhancementLayer}
		{
			\renewcommand{\baselinestretch}{1}\small
	\scalebox{0.72}{%
			\begin{tabular}{|c|c|c|c|c|c|}
				\cline{1-6}
				\multicolumn{1}{|c|}{Layers}&
				\multicolumn{1}{c|}{C1/4/7}&
				\multicolumn{1}{c|}{C2/5/8} &				
				\multicolumn{1}{c|}{C3/6/9}&
				\multicolumn{1}{c|}{C10-14}&
				\multicolumn{1}{c|}{C15}\\
				\cline{1-6}
				\hline
				Filter~size &$3\times3$&$5\times5$&$7\times7$&$3\times3$& $3\times3$\\
				\cline{1-6}
				Filter~number & 32 & 32 & 32 & 32 & 1\\
				\cline{1-6}
				Stride & 1 & 1 & 1 & 1 & 1 \\
				\cline{1-6}
				Function & BN+PReLU & BN+PReLU & BN+PReLU & BN+PReLU & BN+PReLU \\
				\cline{1-6}
			\end{tabular}}}
	\end{center} 
	\vspace{-6mm}
\end{table}

We denote the composite non-linear mapping as $H_{l}(.)$, including Convolution (Conv), PReLU and BN. We further denote the output of the l-th layer as $x_{l}$, such that each layer can be formulated as follows
\begin{equation}
\label{eq:DenseNet}
\begin{split}
x_{11}& = H_{11}([x_{10}]), \\
x_{12}& = H_{12}([x_{10}, x_{11}]), \\
x_{13}& = H_{13}([x_{10}, x_{11}, x_{12}]), \\
x_{14}& = H_{14}([x_{10}, x_{11}, x_{12}, x_{13}]), 
\end{split}
\end{equation} 
where ${x_{10}, x_{11}, x_{12}, x_{13}}$ refers to the concatenation of the feature maps produced in layers C10-C14. Finally, the enhanced target view $\hat{V}_{t}$ is generated by the pixel-wise summation of learned enhancement residual $R_{t}(\theta_{qe})$ and input target view $\tilde{V}_{t}$
\begin{equation}
\label{eq:EnhancedTargetView}
\hat{V}_{t} = \tilde{V}_{t} + R_{t} \left ( \theta_{qe} \right ), \\
\end{equation}
with $\theta_{qe}$ is defined as the trainable parameters of the \ac{mv-qenn}.
The \ac{mv-qenn} is trained with minimizing a mean squared error loss function 
\begin{equation}
\label{eq:LossFi=unctionEN}
\textit{loss} = \|  V_{t} - \hat{V}_{t} \|_{2}^{2}. \\
\end{equation}
It should be noted that the $N_R$ reference views are not enhanced by the \ac{mv-qenn} and thus $\hat{V}_{p_i} = \tilde{V}_{p_i},$ \hbox{$\forall i \in \{1, \dots, N_R \}$.}      



\section{Experimental Results}
\label{sec:experimental}
In this section, we first give the test material used to train the learning-based models and the testing conditions used to assess and compare the proposed solution with respect to state-of-the-art methods. The performance of the proposed solution are then assessed in terms of coding efficiency, visual quality and complexity at both encoder and decoder sides.   
\subsection{Experimental configurations}
\subsubsection{\Ac{lf-d2gan} training}
the proposed \ac{lf-d2gan} architecture described in the previous section was trained with 140 \ac{llf} images, where 70 \ac{llf} images are from EPFL dataset~\cite{Rerabek:218363}, 50 \ac{llf} images are from Stanford Lytro \ac{lf} image dataset~\cite{stanford2018} and 20 \ac{llf} images are from HCI dataset~\cite{Honauer-etal-2017-ICCV}. A validation set was also considered with 14 \ac{llf} images from these three data sets (8, 4 and 2 images from EPFL, Stanford Lytro and HCI datasets, respectively). Each subaperture view was split into patches of size $60\times60$, thus resulting in more than 150,000 patches that were used in the training phase. The training configuration of \ac{lf-d2gan} was set as follows: we trained the generator $G$ and two discriminators ($D_{1}$ and $D_{2}$) with the Adam optimizer~\cite{kingma2014adam} by setting $\beta_{1} = 0.9$, $\beta_{2}= 0.999$, learning rate $= 0.0002$, batch-size of $10$ and kernel size of convolutional layers as depicted in Fig.~\ref{fig:D2GANArchitecture}. The regularization coefficients of $D_{1}$ and $D_{2}$ were set as $\alpha= 0.2$ and $\beta= 0.2$, respectively. For the generator, we used input patch of 60$\times$60, stride of $16$, and output patch equal to $36\times36$ (reduced size is due to the convolutions).
\subsubsection{\Ac{mv-qenn} training} 
the same training data set used to train the \ac{lf-d2gan} was considered to train the \ac{mv-qenn}.  The training set includes both original and decoded views at different \acp{qp}. The views were segmented into patches of \hbox{$64\times64$} as the training samples. The batch size was set to 128 and Adam optimizer~\cite{kingma2014adam} was used with an initial learning rate of $0.0002$. The configurations of the different layers are summarised in Table~\ref{tab:QualityEnhancementLayer}.   
\subsubsection{Testing conditions}
for the testing phase, 9 \ac{llf} images different from the training and validation sets are selected, 6  \ac{llf} images are from EPFL dataset~\cite{Rerabek:218363}, 1 \ac{llf} image from Stanford Lytro \ac{lf} dataset~\cite{stanford2018} and 2 \ac{llf} images from HCI dataset~\cite{Honauer-etal-2017-ICCV}. Each of these \ac{llf} images is composed of 8$\times$8 subaperture views ($N=64$). 
\begin{figure}[t!]
	\centerline{\includegraphics[width=0.5\textwidth]{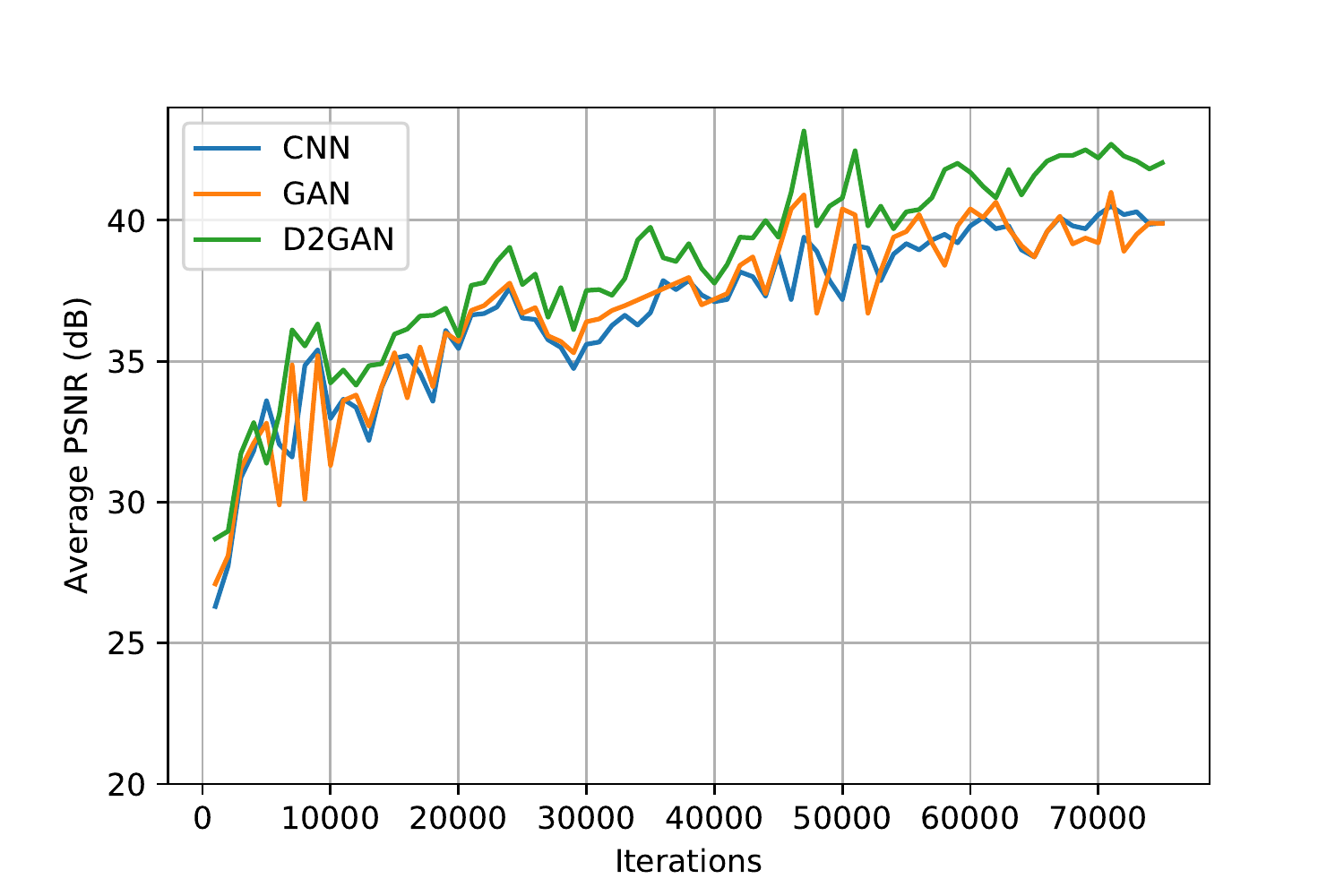}}
	\caption{Average PSNR performance during the training iterations on the validation set for \ac{cnn}, \ac{gan} and the proposed \ac{lf} non-reference views synthesis (\acs{lf-d2gan}) architectures.}
	\label{fig:Comparaison}
\end{figure}

The \ac{lf} image views are arranged in a pseudo-video sequence using spiral order scan and encoded using \ac{vvc} in \ac{ra} coding configuration at 4 \ac{qp} values of 18, 24, 28 and 32. The \ac{vtm} version 7.1 is used to encode the pseudo-video sequence in YCbCr 4:2:0 sampling color format. The $N_R=16$ reference views are selected as the four corner views of each quadrant as illustrated in Fig.~\ref{fig:sparse-rep}. In this figure, the central reference view is highlighted in red color while the rest of disabled 48 views correspond to the non-reference views.  
\begin{figure*}[t!]
    \subfloat[\label{fig:anchor_qp17_edgedetection}{\it Bikes}]{\includegraphics[width=0.33\textwidth]{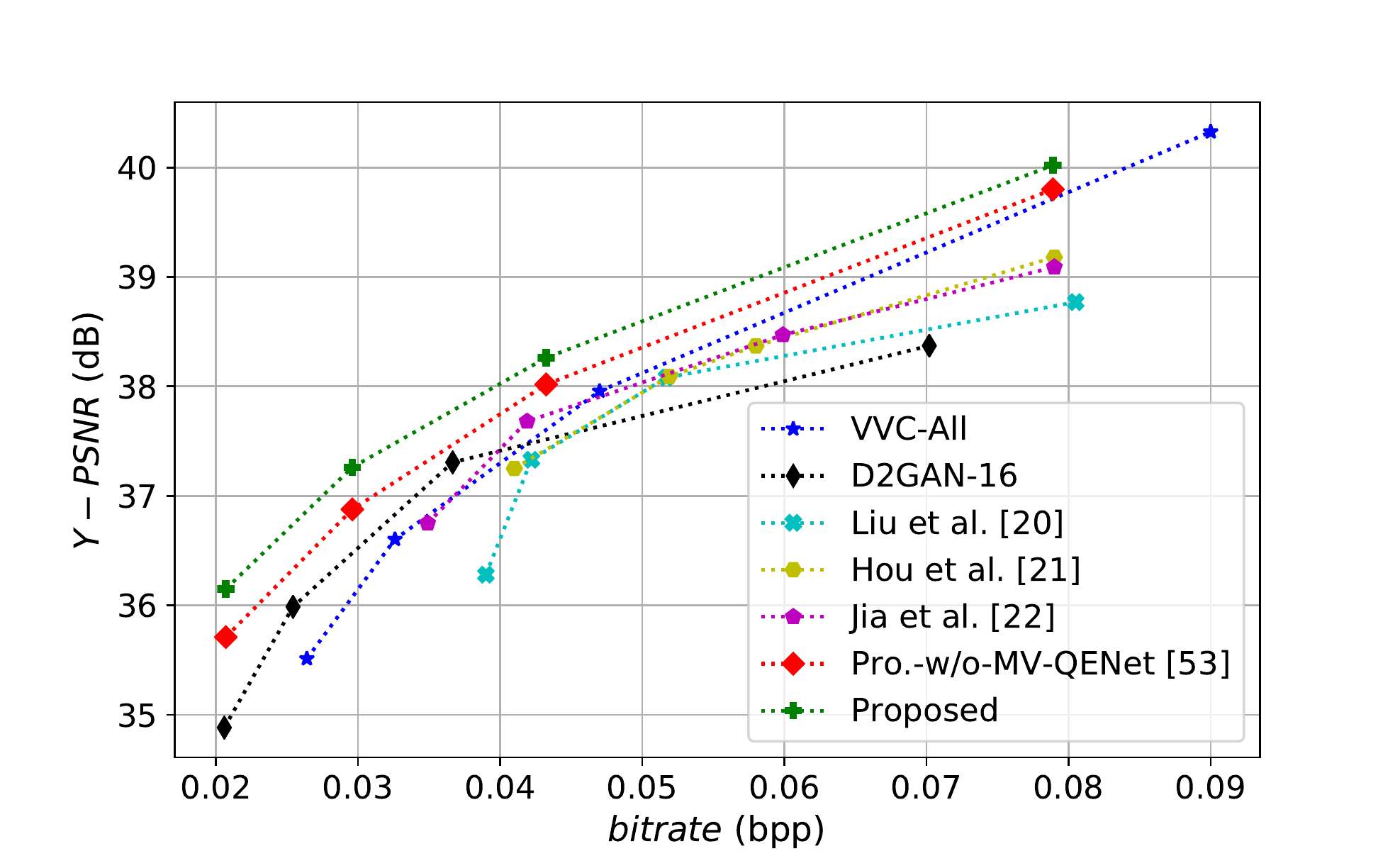}}    
	\subfloat[\label{fig:anchor_qp17_edgedetection}{\it Danger de Mort}]{\includegraphics[width=0.33\textwidth]{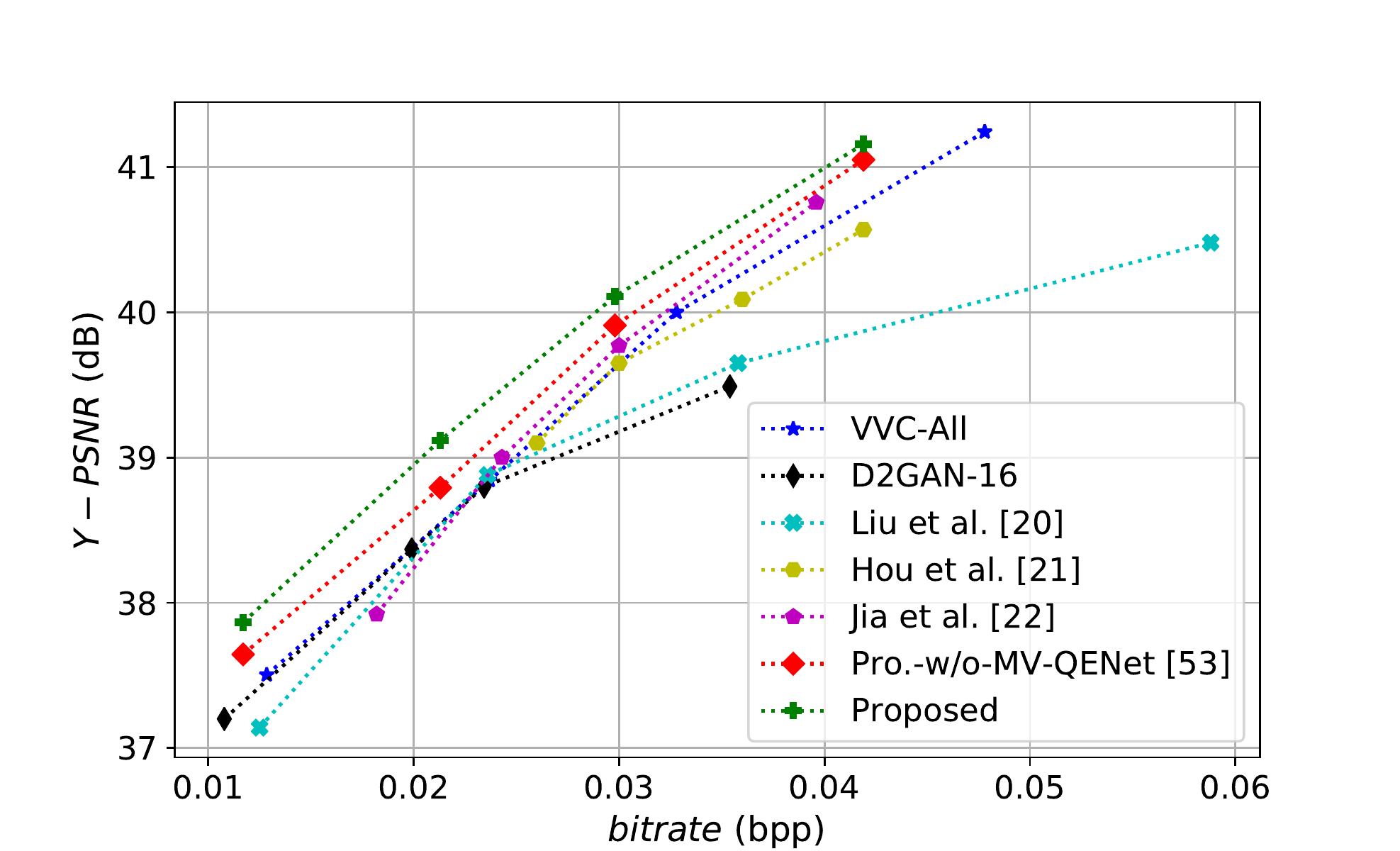}} 
	\subfloat[\label{fig:anchor_qp17_edgedetection}{\it Flowers}]{\includegraphics[width=0.33\textwidth]{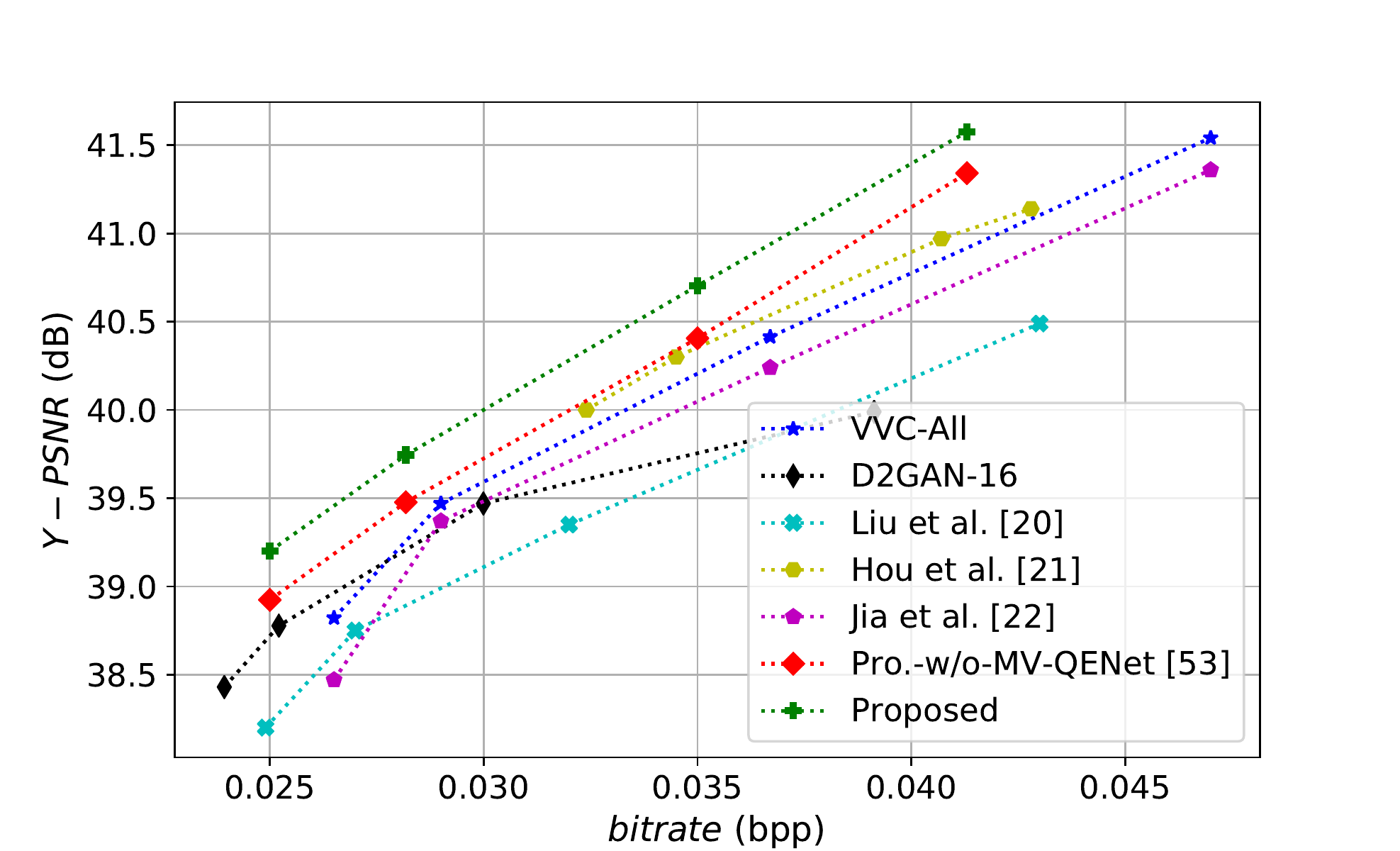}}\vspace{-0.35cm} 
	
	\subfloat[\label{fig:anchor_qp17_edgedetection}{\it Ankylosaurus Diplodocus}]{\includegraphics[width=0.33\textwidth]{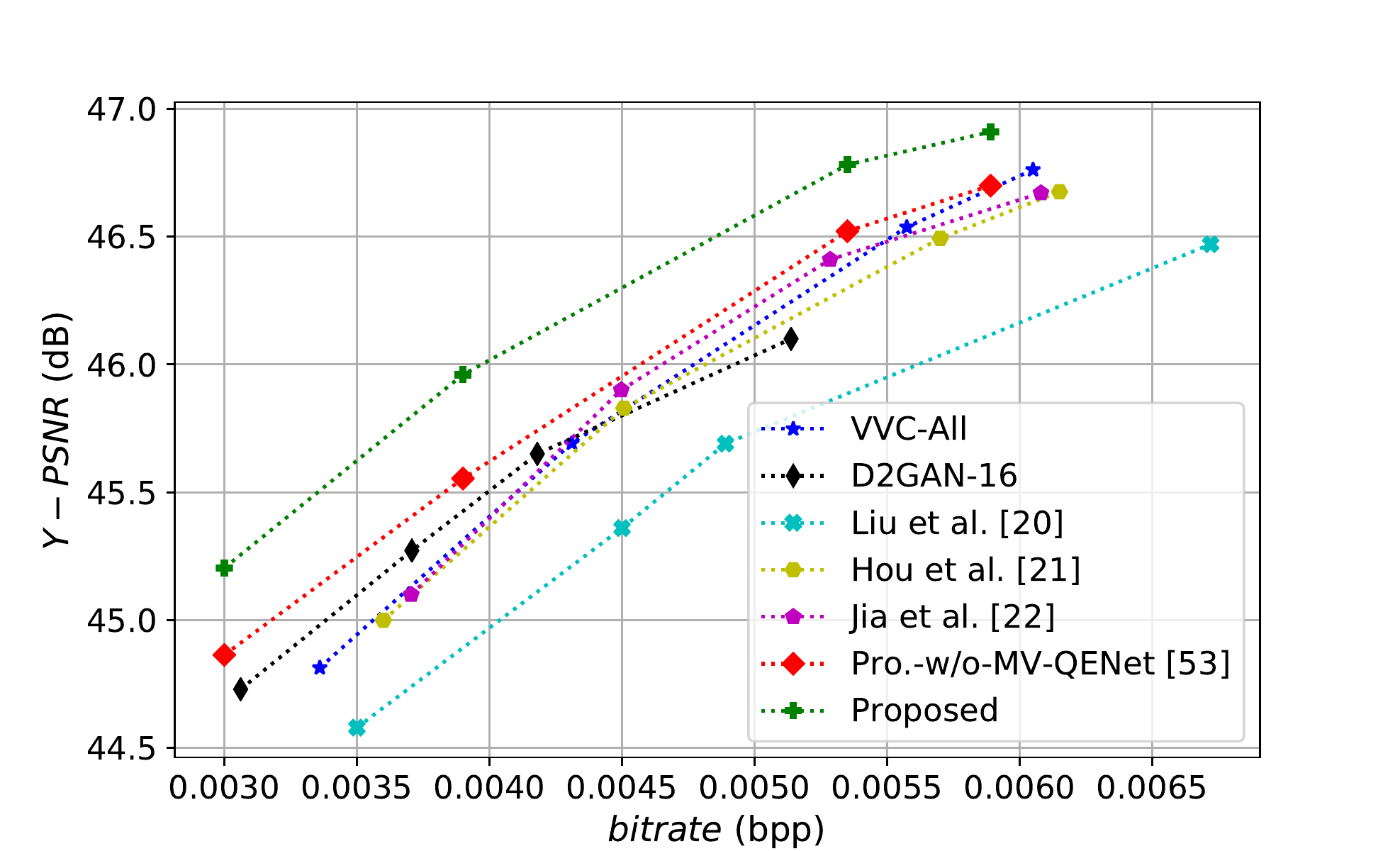}}     \hfill
	\subfloat[\label{fig:anchor_qp17_edgedetection}{\it Aloe}]{\includegraphics[width=0.33\textwidth]{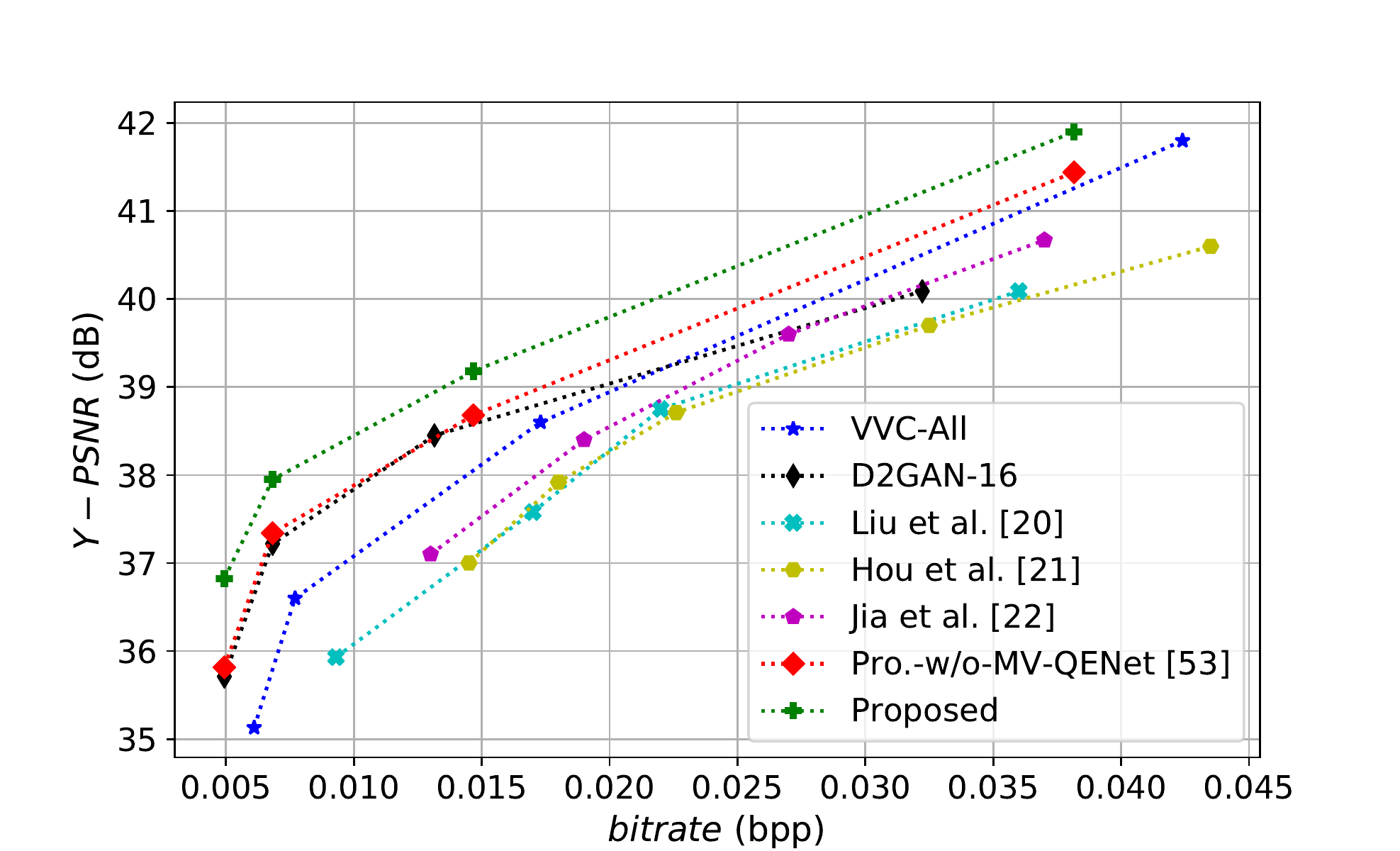}}     \hfill
	\subfloat[\label{fig:anchor_qp17_edgedetection}{\it Stone-Pillars-Outside}]{\includegraphics[width=0.33\textwidth]{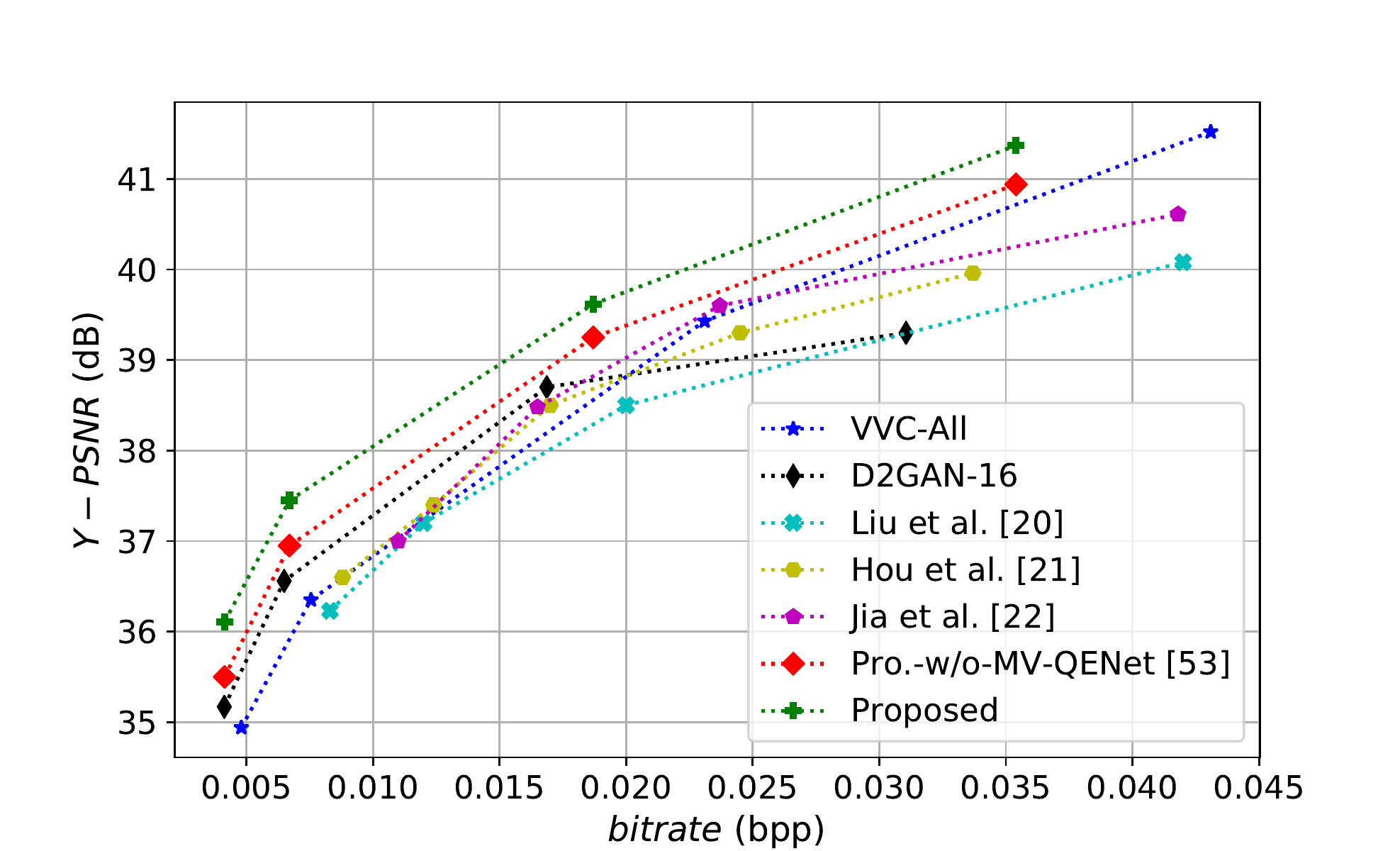}}\vspace{-0.35cm} 
	
    \subfloat[\label{fig:anchor_qp17_edgedetection}{\it Bedroom}]{\includegraphics[width=0.33\textwidth]{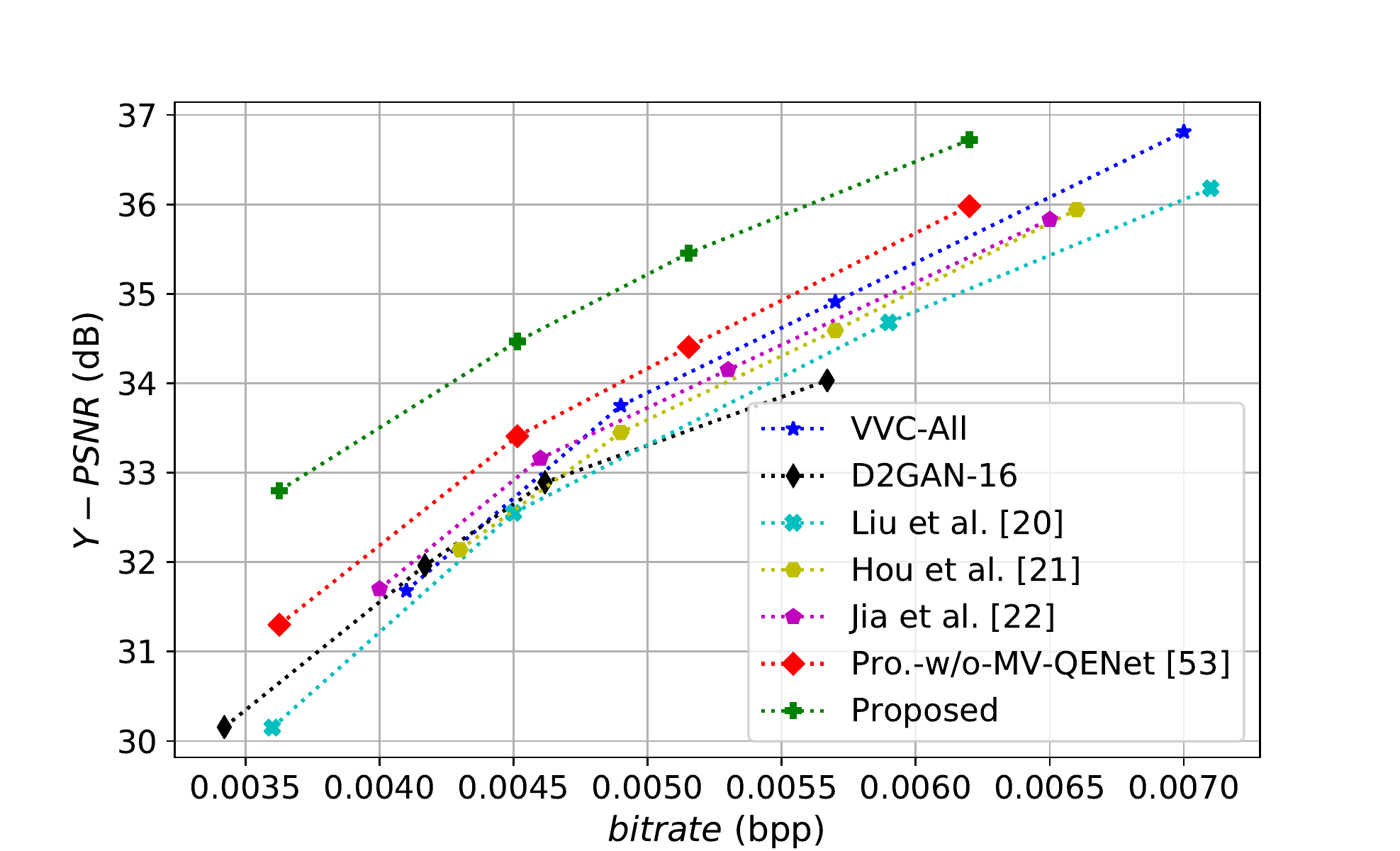}}     \hfill
    \subfloat[\label{fig:anchor_qp17_edgedetection}{\it Desktop}]{\includegraphics[width=0.33\textwidth]{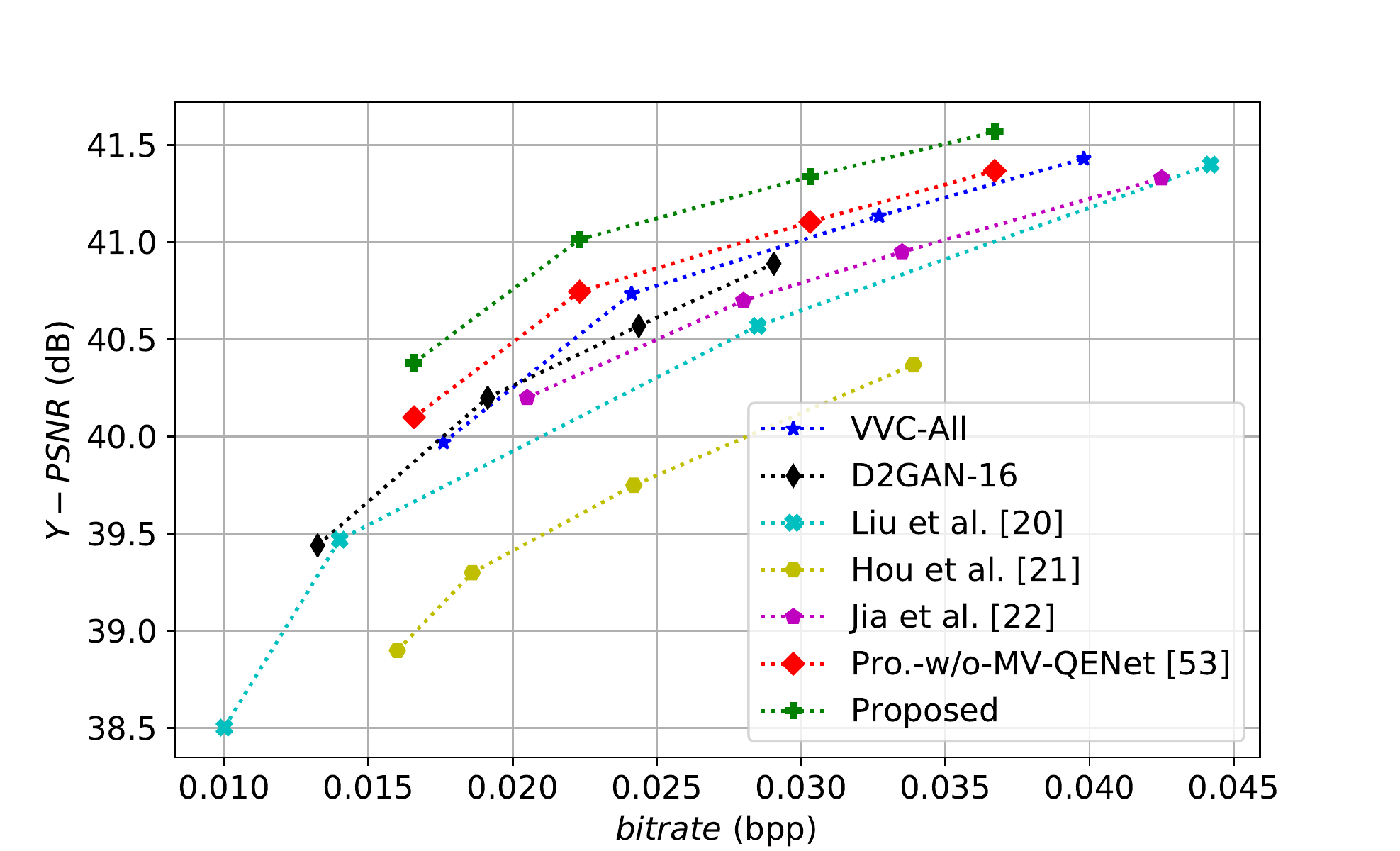}}     \hfill
    \subfloat[\label{fig:anchor_qp17_edgedetection}{\it Herbs}]{\includegraphics[width=0.33\textwidth]{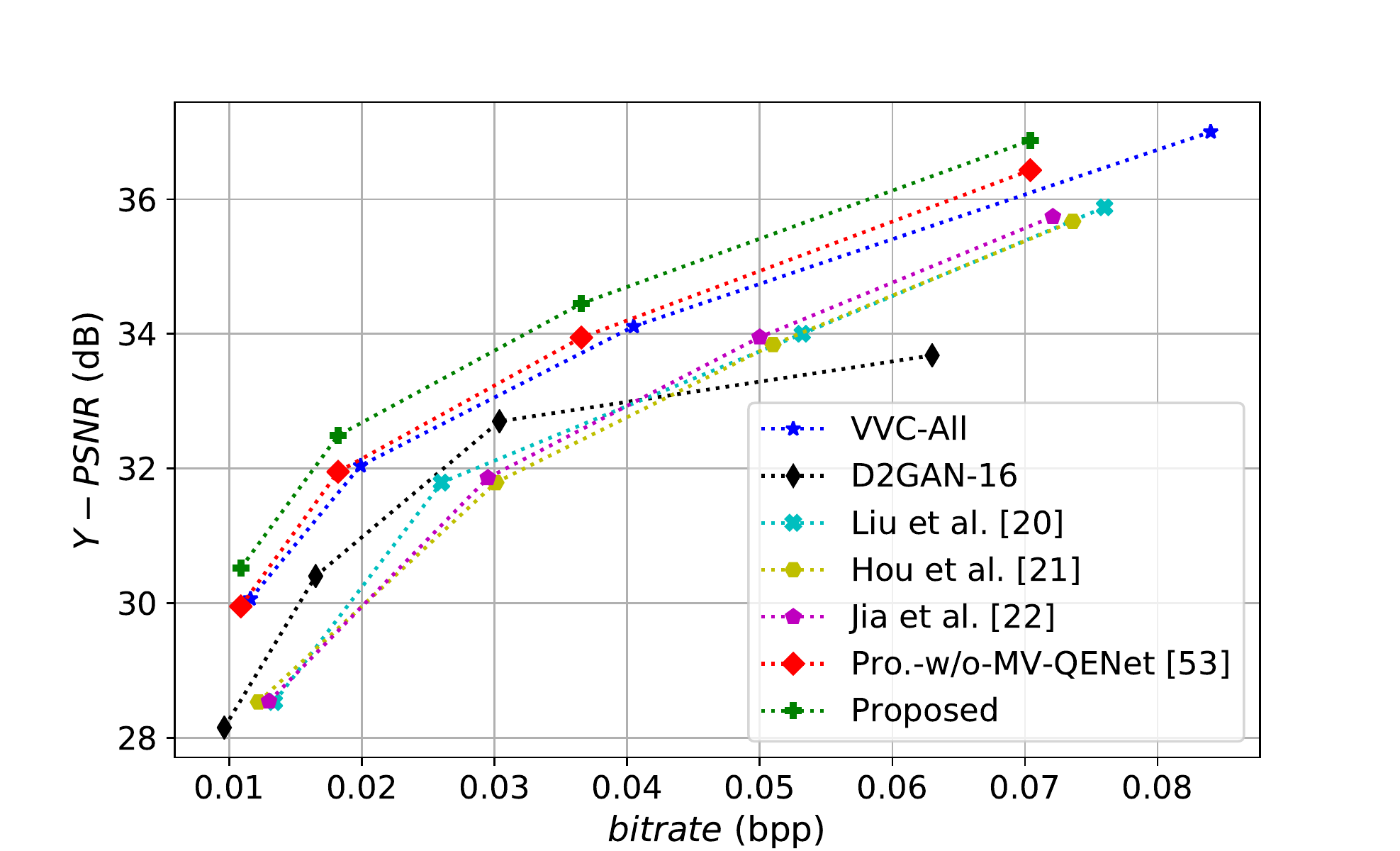}}     \hfill
	\caption{RD curves of the five considered solutions for the 9 \ac{lf} images using four QP values.}
	\label{fig:results_psnr}
	\vspace{-3mm}
\end{figure*}

The proposed solution is compared with respect to six coding solutions including 1) \ac{vvc}-All that encodes all views with \ac{vvc} standard, 2) \ac{lf-d2gan}-16 that encodes the 16 reference views with \ac{vvc} and the non-reference views are synthesized with the \ac{lf-d2gan}, 3)  Liu {\it et al.} method~\cite{7574674}, 4) Hou {\it et al.} method~\cite{Hou:2019:LFI:3347960.3347993}, 5) Jia {\it et al.} method~\cite{8574895} and \hbox{6) the} proposed solution without the quality enhancement block (denoted as  prop.-w/o-\ac{mv-qenn})~\cite{9102880}. The quality of the decoded views is assessed using both \ac{psnr} and \ac{ssim} \cite{wang2004image} \ac{iqa} metrics.

\subsection{Coding and quality evaluation performance}
Fig.~\ref{fig:Comparaison} illustrates the average \ac{psnr} versus training iterations of the synthesis block on the validation set for three different architectures: \ac{cnn}, \ac{gan} and the proposed \ac{lf-d2gan}. The \ac{cnn} architecture is trained by minimizing the mean squared-error loss function between the synthesized and the original views, while the \ac{gan} and \ac{lf-d2gan} architectures are trained with one and two adversarial discriminators, respectively. It is clear from this figure that the proposed \ac{lf-d2gan} architecture relying on two discriminators provides higher \ac{psnr} quality performance on the validation set with smooth fluctuations of the quality and better convergence of the generator compared to both \ac{cnn} and \ac{gan} architectures.      

Fig.~\ref{fig:results_psnr} gives the average \ac{psnr} performance versus the bitrate for the proposed and the six considered reference solutions on the 9 testing \ac{lf} images.  The first important observation is that the proposed solution performs better than the six reference solutions at all considered bitrates and for all test \ac{lf} images. 
We can also notice that the three proposed components performing view synthesis (\ac{lf-d2gan}), \ac{rd} optimization and quality enhancement (\ac{mv-qenn}) bring significant quality improvements since our solution performs better than \ac{vvc}-All, \ac{lf-d2gan}-16 and prop.-w/o-\ac{mv-qenn} solutions. 

Table~\ref{tab:BD} gives the performance of our solution and four reference solutions in terms of \ac{bd-br} and \ac{bd-psnr}, both computed with respect to the anchor solution proposed by Liu {\it et al.} in~\cite{7574674}. We can notice that our solution achieves the highest \ac{bd-br} and \ac{bd-psnr} gains for the 9 test \ac{lf} images. In average, our solution provides 36.22\% bitrate reduction and increases the quality by 1.35 dB compared to the solution proposed in~\cite{7574674}. Compared to the second best performing solution, i.e., prop.-w/o-\ac{mv-qenn}~\cite{9102880}, our solution offers a relative bitrate gain of 8.12\% and increases the quality by 0.52 dB. These scores highlight the significant gains brought by the different proposed blocks in terms of bitrate reduction and quality enhancement. These average gains are also substantial for the 9 individual test \ac{lf} images.

\begin{table*}[!h]
\renewcommand{\arraystretch}{1.1} 
	\begin{center}
		\caption{\acs{bd-br} and \acs{bd-psnr} performance calculated with respect to the anchor method proposed in~\cite{7574674}.}
		\label{tab:BD}
		\footnotesize
		\begin{tabular}{|l|c|c|c|c|c|c|c|c|c|c|}
			\cline{1-11}
			\multicolumn{1}{|c|}{~}&
			\multicolumn{2}{c|}{\ac{vvc}-All}&
			\multicolumn{2}{c|}{Jia \textit{et al.}~\cite{8574895}}&
			\multicolumn{2}{c|}{Hou \textit{et al.}~\cite{Hou:2019:LFI:3347960.3347993}}&
			\multicolumn{2}{c|}{Prop.-w/o-\ac{mv-qenn}~\cite{9102880}}&
			\multicolumn{2}{c|}{Proposed}\\
			\cline{2-11}
			\multicolumn{1}{|c|}{Image} &
			\acs{bd-br}& \acs{bd-psnr} & \acs{bd-br}& \acs{bd-psnr} & \acs{bd-br} & \acs{bd-psnr} &\acs{bd-br}&\acs{bd-psnr}&\acs{bd-br}&\acs{bd-psnr}\\
			\hline
			\textit{Bikes} &--11.7\% &0.72  &--6.3\% &0.48 &--6.9\% &0.49  &--22.4\% &0.96&\textbf{--31.56}\% &\textbf{1.19} \\			
			\textit{Danger De Mort} &--7.8\% &0.22 &--10.8\% &0.28 &--8.7\% &0.26 &--16.5\% &0.40 &\textbf{--25.69}\% &\textbf{0.78}\\
			\textit{Flowers} &--12.3\% &0.56 &--11.9\% &0.54 &--16.2\% &0.72 &--16.6\% &0.74 &\textbf{--23.66}\% &\textbf{1.03}\\				
			\textit{Ankylosaurus Dip1} &--13.2\% &0.44 &--14.9\% &--0.72 &--12.3\% &0.39 &--18.0\% &0.57 &\textbf{--31.17}\% &\textbf{1.15}\\				
			\textit{Aloe} &--26.4\% &0.85 &--9.1\% &0.31 &--2.46\% &--0.12 &--42.3\% &1.23 &\textbf{--56.59}\% &\textbf{1.84}\\				
			\textit{Stone-pillars-outside} &--18.3\% &0.61 &--15.1\% &0.52 &--11.9\% &0.28 &--35.6\% &0.98 &\textbf{--49.76}\% &\textbf{1.42}\\				
			\textit{Bedroom} &--5.3\% &0.46 &--4.0\% &0.32 &--2.3\% &0.18 &--9.5\% &0.85 &\textbf{--24.78}\% &\textbf{2.11}\\				
			\textit{Desktop} &--19.6\% &0.32 &--7.5\% &0.11 &44.1\% &--0.61 &--26.3\% &0.45 &\textbf{--40.58}\% &\textbf{0.79}\\				
			\textit{Herbs} &--26.0\% &1.14 &--4.4\% &--0.11 &6.9\% &--0.20 &--29.8\% &1.32 &\textbf{--42.25}\% &\textbf{1.85}\\				
			\cline{1-11}				
			Average  &--15.6\% &0.59 &--8.3\% &0.35  &--0.54\% &0.15 &--24.1\% &0.83 &\textbf{--36.22}\% &\textbf{1.35} \\							
			\hline
		\end{tabular}
		\vspace{-4mm}
	\end{center}
\end{table*}

\begin{figure*}[t!]
 \subfloat[\label{fig:anchor_qp17_edgedetection}{QP  18}]{\includegraphics[width=0.5\textwidth]{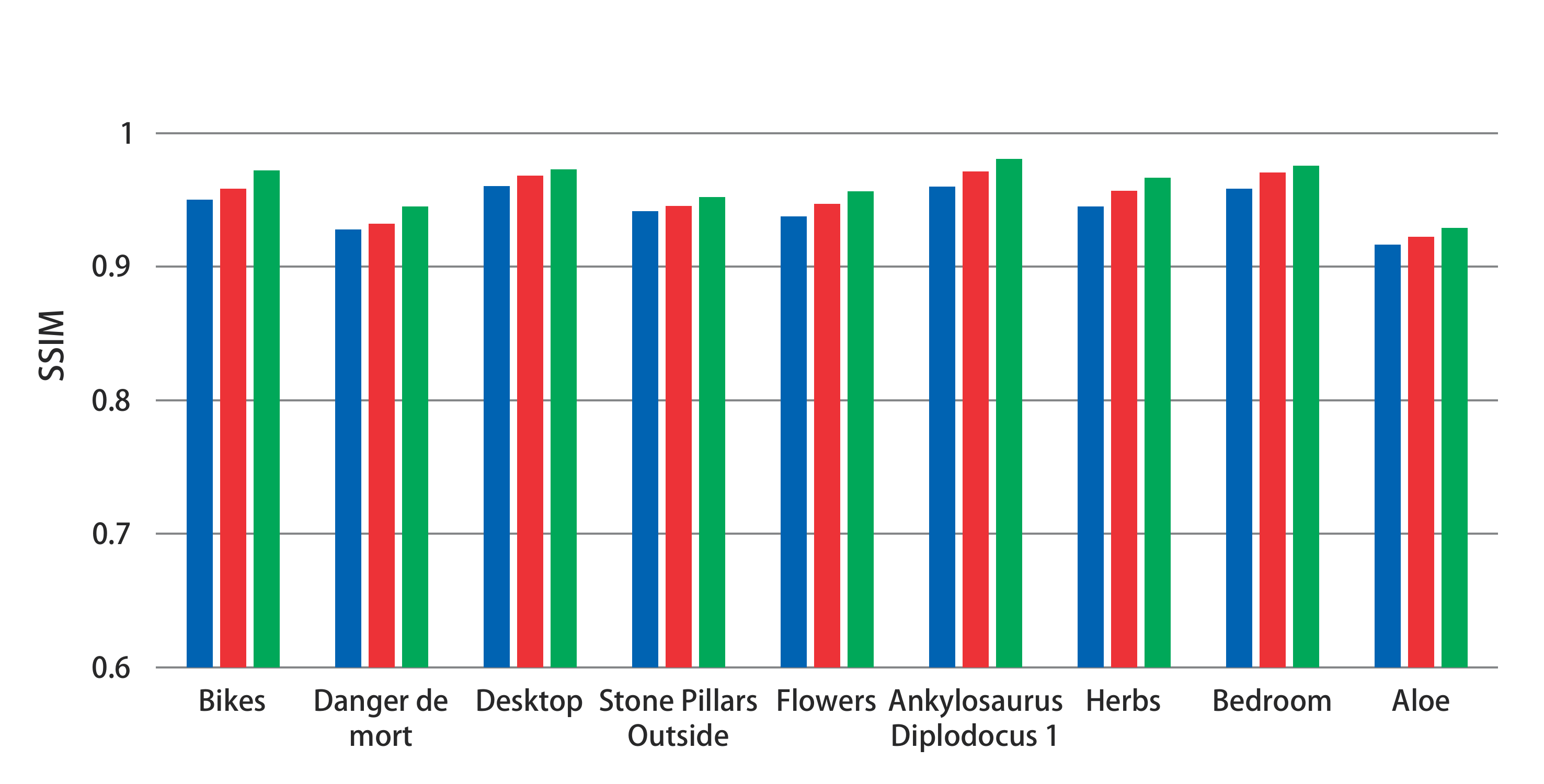}}  
 \subfloat[\label{fig:anchor_qp17_edgedetection}{QP  32}]{\includegraphics[width=0.5\textwidth]{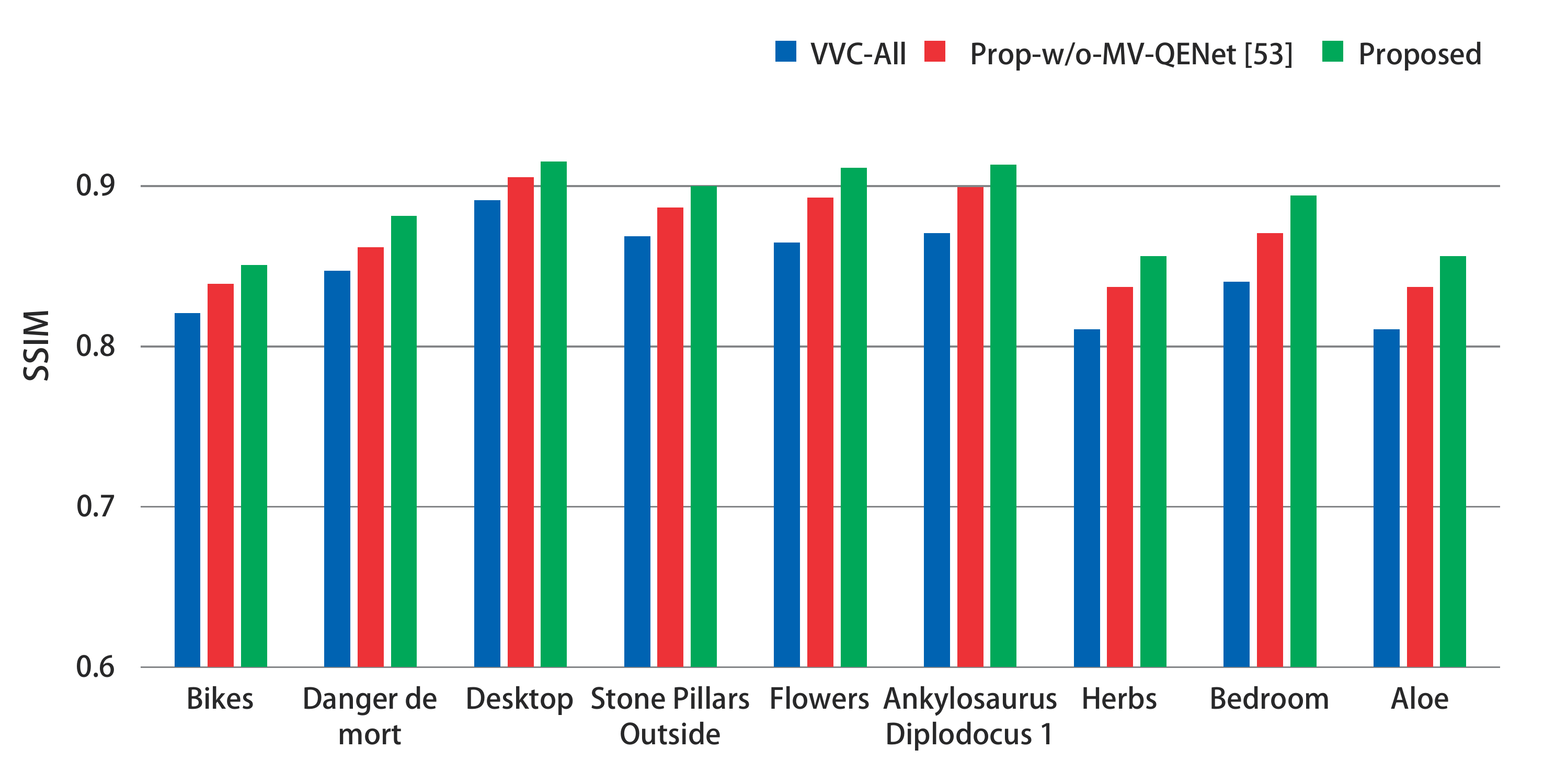}}  
 
	\caption{\acs{ssim} performance of three \ac{lf} image coding methods for the 9 considered test \acs{lf} images at two \ac{qp} values.}
	\label{fig:SSIM}
		\vspace{-2mm}
\end{figure*}

\begin{figure*}[!h]
	\centerline{\includegraphics[width=0.97\textwidth]{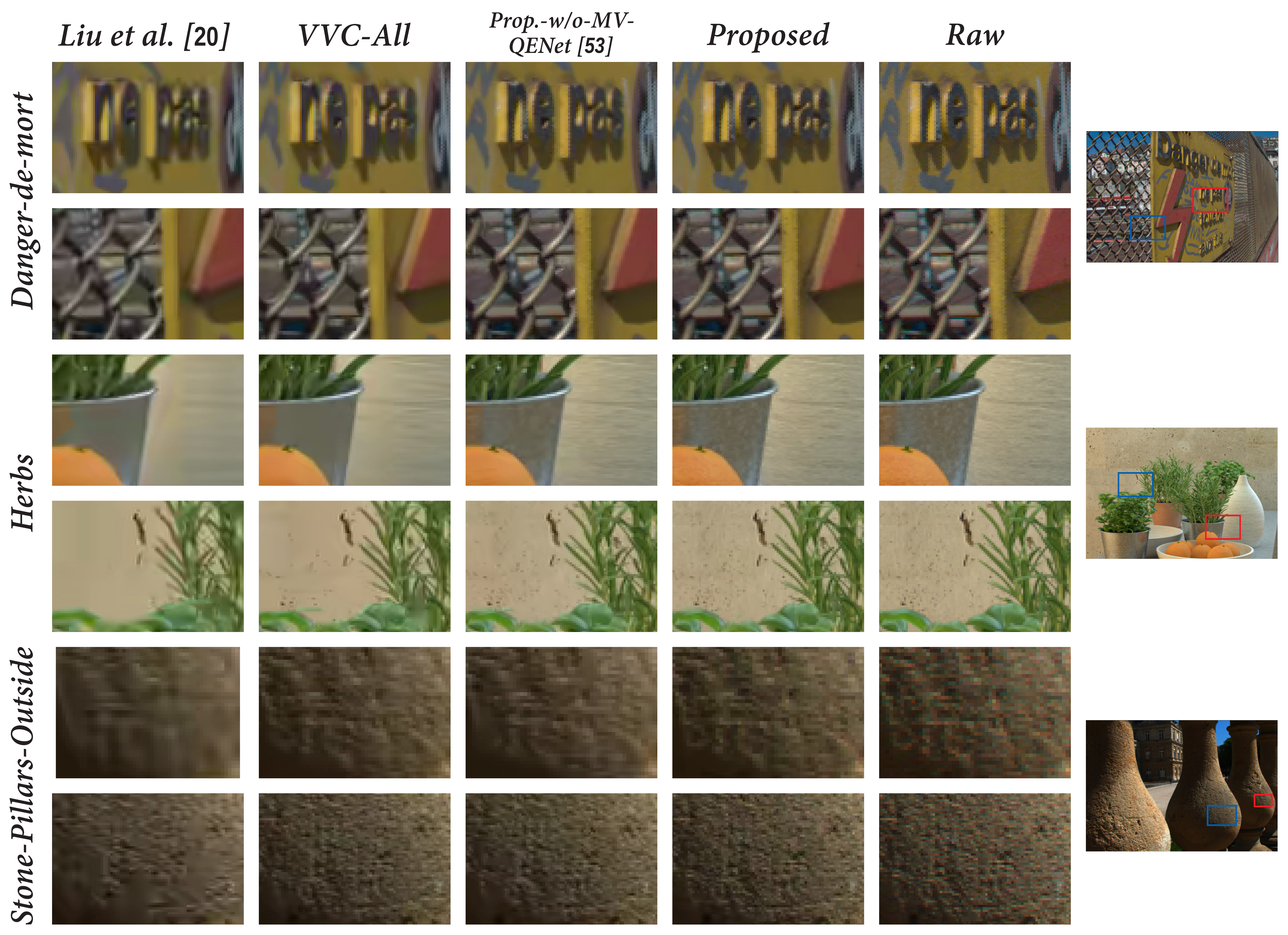}}
	\caption{\small {Visual illustration of 3 test \ac{lf} images encoded at around 0.014 bpp for {\it Danger de Mort}, 0.02 bpp for {\it Herbs} and 0.01 bpp for {\it Stone-Pillars-Outside}. The objective quality scores are provided for each illustrated view in this format PSNR(SSIM). {\it Danger de Mort} [view 49 (TL $\#4$), \ac{rdo}: Synthesized by \ac{lf-d2gan}], Liu {\it et al.} 37.0(0.84), \ac{vvc}-All 37.2(0.87), prop.-w/o-\ac{mv-qenn} 37.4(0.88), proposed 37.8(0.91); {\it Herbs} [view 47 (TL $\#4$), \ac{rdo}: Synthesized by \ac{lf-d2gan}], Liu {\it et al.} 30.8(0.80), \ac{vvc}-All 31.7(0.85), prop.-w/o-\ac{mv-qenn} 32.1(0.87), Proposed 32.6(0.90) and {\it Stone-Pillars-Outside}, [view 33 (TL $\#4$), \ac{rdo}: decoded by \ac{vvc}], Liu {\it et al.} 35.7(0.81), \ac{vvc}-All 36.2(0.84), prop.-w/o-\ac{mv-qenn} 36.2(0.84), prop. 36.7(0.87). }}
	\label{fig:VisualQuality}
	\vspace{-2mm}
\end{figure*}

Fig.~\ref{fig:SSIM} gives the \ac{ssim} performance of the \ac{vvc}-All, prop.-w/o-\ac{mv-qenn} and proposed methods for the 9 test \ac{lf} images at two \ac{qp} values 18 and 32.  We can notice from this figure that our solution gives the highest \ac{ssim} scores for all \ac{lf} images at both considered \acp{qp}. Table~\ref{tab:BDSSIM} shows the \ac{ssim}-based \ac{bd-br} and \ac{bd-ssim} of our solution and two other methods with respect to the anchor method proposed in~\cite{7574674}. Our solution achieves the highest bitrate saving in average with around $42.95\%$ compared to~\cite{7574674}. These scores are even higher compared to the \ac{psnr}-based birate savings reported in Table~\ref{tab:BD}.  Compared to the prop.-w/o-\ac{mv-qenn} solution~\cite{9102880}, we can notice a relative bitrate saving of $12.94\%$ in average, which highlights the contribution of the proposed \ac{mv-qenn} post-processing.     

\begin{table}[!h]
\renewcommand{\arraystretch}{1.2} 
	\begin{center}
		\caption{\ac{ssim}-based \acs{bd-br} and \acs{bd-ssim} performance calculated with respect to the anchor~\cite{7574674}. 1) \textit{Bikes}, 2) \textit{Danger de Mort}, 3) \textit{Flowers},  4) \textit{Ankylosaurus Dip1}, 5) \textit{Aloe} , 6) \textit{Stone-pillars-outside}, 7) \textit{Bedroom}, 8) \textit{Desktop}, 9) \textit{Herbs}.}
		\label{tab:BDSSIM}
	
		        \resizebox{\linewidth}{!}{
        \begin{scriptsize}
		\begin{tabular}{|l|c|c|c|c|c|c|}
			\cline{1-7}
			\multicolumn{1}{|c|}{~}&
			\multicolumn{2}{c|}{\ac{vvc}-All}&
			\multicolumn{2}{c|}{Prop.-w/o-\ac{mv-qenn}~\cite{9102880}}&
			\multicolumn{2}{c|}{Proposed}\\
			\cline{2-7}
			\multicolumn{1}{|c|}{Im.} &
			\acs{bd-br}&\acs{bd-ssim}&\acs{bd-br}&\acs{bd-ssim}&\acs{bd-br}&\acs{bd-ssim}\\
			\hline
			1) &--10.3\%&0.011 &--24.91\% &0.022 &{\bf --40.1\%} &{\bf 0.032}  \\			
			2) &--12.4\%&0.013 &--22.10\% &0.021 &{\bf--33.7\%} &{\bf 0.027}  \\
			3) &--13.0\%&0.020 &--23.15\% &0.028 &{\bf--32.8\%} &{\bf 0.042}  \\				
			4) &--16.01\%&0.011 &--22.8\% &0.024 &{\bf--40.0\%} &{\bf 0.035}  \\				
			5) &--30.0\%&0.028 &--50.30\% &0.046 &{\bf--66.6\%} &{\bf 0.049}  \\				
			6) &--19.80\%&0.004 &--40.49\% &0.021 &{\bf--54.0\%} &{\bf 0.031}  \\				
			7) &--9.4\%&0.010 &--17.22\% &0.015 &{\bf--30.6\%} &{\bf 0.024}  \\				
			8) &--18.0\%&0.011 &--37.09\% &0.029 &{\bf--42.5\%} &{\bf 0.036}  \\				
			9) &--19.8\%&0.020 &--32.10\% &0.041 &{\bf--46.1\%} &{\bf 0.059}  \\				
			\cline{1-7}				
			Av.  &--16.54\%&0.014 &--30.01\% &0.027 &{\bf--42.95\%} & {\bf 0.037}  \\		
			\hline
		\end{tabular}
        \end{scriptsize}}
	\end{center}
		\vspace{-2mm}
\end{table}

Fig.~\ref{fig:VisualQuality} illustrates the visual quality of the decoded non-reference views resulting from Liu {\it et al.}\cite{7574674}, \ac{vvc}-All, prop.-w/o-\ac{mv-qenn} and proposed methods for 3 test \ac{lf} images: {\it Danger de Mort}, {\it Herbs} and {\it Stone-Pillars-Outside}. We can see that our method provides a higher visual quality of the reconstructed views, especially after applying the \ac{mv-qenn} post-processing. This latter enhances the visual quality of views by providing more details (high frequencies), especially at the edges.  

\subsection{Complexity analysis}
The complexity of the proposed coding approach is evaluated and compared to the other methods on both CPU and GPU platforms. The performance has been carried-out on a PC equipped with an Intel core i9-7900X CPU running at 3.3 GHz with 64 GB memory and a TITAN Xp NVDIA GPU. The complexity of our solution is assessed on  CPU, where all modules run on the CPU, and on GPU when both \ac{lf-d2gan} and \ac{mv-qenn} modules run on the GPU. Table~\ref{tab:TimeComplexity} gives the encoding and decoding times in second for our solution and three other methods including \ac{vvc}-All, Jia {\it et al.}~\cite{8574895} and Liu {\it et al.}~\cite{7574674} methods. The complexity of the proposed encoder is in the same range as the complexity of the solution proposed in \cite{8574895} that also relies on a \ac{gan} to synthesize the non-reference views at the encoder. We can also notice that the GPU enables to speedup the \ac{lf-d2gan} and \ac{mv-qenn} blocks at both encoder and decoder. The complexity of the proposed encoder is in average 6$\times$ faster than the encoder proposed in~\cite{7574674}. This latter relies on the \ac{jem} codec which is  more complex than the \ac{vtm} codec. However, the proposed decoder is more complex than the other decoders. On average, the \ac{vvc} decoding takes 3 seconds, view synthesis using \ac{lf-d2gan} 92 seconds and finally \ac{mv-qenn} block 190 seconds on GPU, which corresponds to 1.05\%, 32.28\% and 66,66\% of the total decoder time, respectively. This clearly shows  that the increase in the complexity of decoder  is mainly due to the synthesis block and in particular to the  quality enhancement block.
\begin{table}[t!]
\renewcommand{\arraystretch}{1.1} 
	\begin{center}
		\caption{Processing time of four \acs{lf} image coding methods.}
		\label{tab:TimeComplexity}
		{
			\renewcommand{\baselinestretch}{1}\footnotesize
			\begin{tabular}{|c|c|c|c||c|c|}
				\cline{1-6}
				\multicolumn{1}{|c|}{~}&
				\multicolumn{5}{c|}{Encoder time in seconds}\\
				\cline{2-6}
				\multicolumn{1}{|c|}{QP}&
				\multicolumn{1}{c|}{VVC-All} &				
				\multicolumn{1}{c|}{Jia \textit{et al.}~\cite{8574895}}&
				\multicolumn{1}{c||}{Liu \textit{et al.}~\cite{7574674}}&
				\multicolumn{2}{c|}{Our}\\
				\cline{2-6}
				\multicolumn{1}{|c|}{~}&
				\multicolumn{1}{c|}{CPU}&
				\multicolumn{1}{c|}{GPU}&
				\multicolumn{1}{c||}{CPU}&
				\multicolumn{1}{c|}{CPU}&
				\multicolumn{1}{c|}{GPU}\\
				\cline{1-6}
				\hline
				18 & \textbf{259} & 450 & 3535 & 559 & 514\\
				\cline{1-6}
				22 & \textbf{152} & 350 & 3030 & 452 & 402\\
				\cline{1-6}
				28 & \textbf{101} & 220 & 2478 & 401 & 349 \\
				\cline{1-6}
				34 & \textbf{66} & 142 & 1710 & 366 & 315 \\
				\cline{1-6}
				Average & \textbf{144} & 291 & 2688 &  445 & 395 \\
				\hline\hline
				\multicolumn{1}{|c|}{~}&
				\multicolumn{5}{c|}{Decoder time in seconds}\\
				\hline
				Average & \textbf{4} & 53 & 5 & 333 & 285\\
				\hline
		\end{tabular}}
	\end{center} 
		\vspace{-6mm}
\end{table}

As we can see in Fig.~\ref{fig:Schema}, the decoder can be optimized by processing several decoding blocks in parallel. In addition, the \ac{mv-qenn} block is optional  and may or may not be applied depending on the computational resources available at the decoder to the detriment of lower quality.    
\section{Conclusion}
\label{sec:colusion}
In this paper, we have proposed an efficient lossy coding scheme for \ac{llf} imaging in subaperture representation. The coding scheme is composed of four elementary blocks, including 2D video coding, view synthesis, rate-distortion optimization and view quality enhancement. The \ac{lf} views are first arranged in a pseudo-video sequence which is encoded with the \ac{vvc} standard in hierarchical temporal scalability configuration. The reference views are encoded at low temporal layers,  while the rest of views are encoded at higher temporal layers. This coding structure enables to drop thanks to \ac{rdo} block the non-reference views without impacting the decoding of reference views. The training of the proposed \ac{lf-d2gan} synthesis block is guided by two adversarial discriminators enabling better convergence of the generator and providing higher \ac{psnr} quality performance of the synthesized views.  A novel quality enhancement block \ac{mv-qenn} is applied at the decoder side on the non-reference views to further enhance their quality and ensure quality consistency between views.

The proposed coding solution has been assessed in terms of bitrate saving and visual quality using both \ac{psnr} and \ac{ssim} objective quality metrics. A significant bitrate saving has been achieved by the proposed method without affecting the visual quality. The obtained results clearly demonstrated the superiority of our solution with respect to the state-of-the-art methods.

As future work, we plan to consider more advanced \ac{lf} image features such as the visual attention and viewing conditions.


%

\ifCLASSOPTIONcaptionsoff
  \newpage
\fi



%


\bibliographystyle{IEEEtran}
\bibliography{IEEEexample}

\begin{IEEEbiography}[{\includegraphics[width=1in,height=1.25in,clip,keepaspectratio]{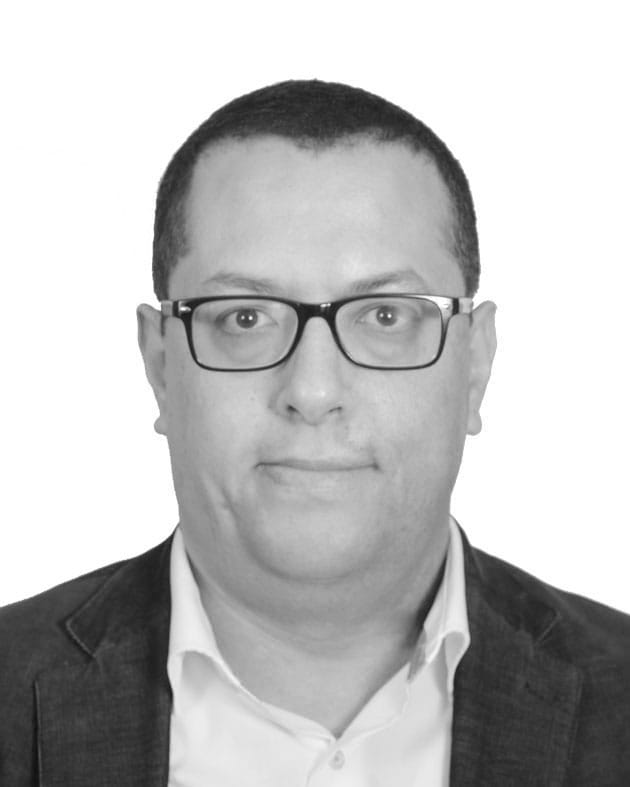}}]{Nader Bakir} received the master’s degree in Applied mathematics from the Faculty of Science, Lebanese University, Beirut, Lebanon, in 2003, and the Ph.D. degree in signal and image processing from National Institute of Applied Sciences of Rennes, Institute of Electronics and Telecommunications of Rennes, Rennes, France, in 2020. Currently, he is an Lecturer at the Faculty of  Business Administration, and researcher in the Center of Research and Studies in Legal Informatics at Lebanese University. His research is in the image processing field.
\end{IEEEbiography}

\begin{IEEEbiography}[{\includegraphics[width=1in,height=1.25in,clip,keepaspectratio]{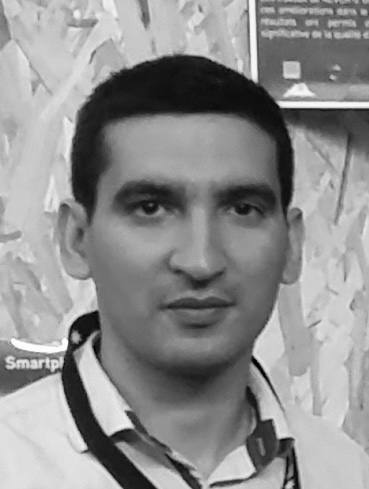}}]{Wassim Hamidouche}
received Master’s and Ph.D. degrees both in Image Processing from the University of Poitiers (France) in 2007 and 2010, respectively. From 2011 to 2013, he was a junior scientist in the video coding team of Canon Research Center in Rennes (France). He was a post-doctoral researcher from Apr. 2013 to Aug. 2015 with VAADER team of IETR where he worked under collaborative project on HEVC video standardisation. Since Sept. 2015 he is  an Associate Professor at INSA Rennes and a member of the VAADER team of IETR Lab. He has joined the Advanced Media Content Lab of b$<>$com IRT Research Institute as an academic member in Sept. 2017. His research interests focus on video coding and multimedia security. He is the author/coauthor of more than one hundred and thirty (+130) papers at top journals and conferences in Image Processing, two MPEG standards, two patents, several MPEG contributions, public datasets and open source software projects.
\end{IEEEbiography}

\begin{IEEEbiography}[{\includegraphics[width=1in,height=1.25in,clip,keepaspectratio]{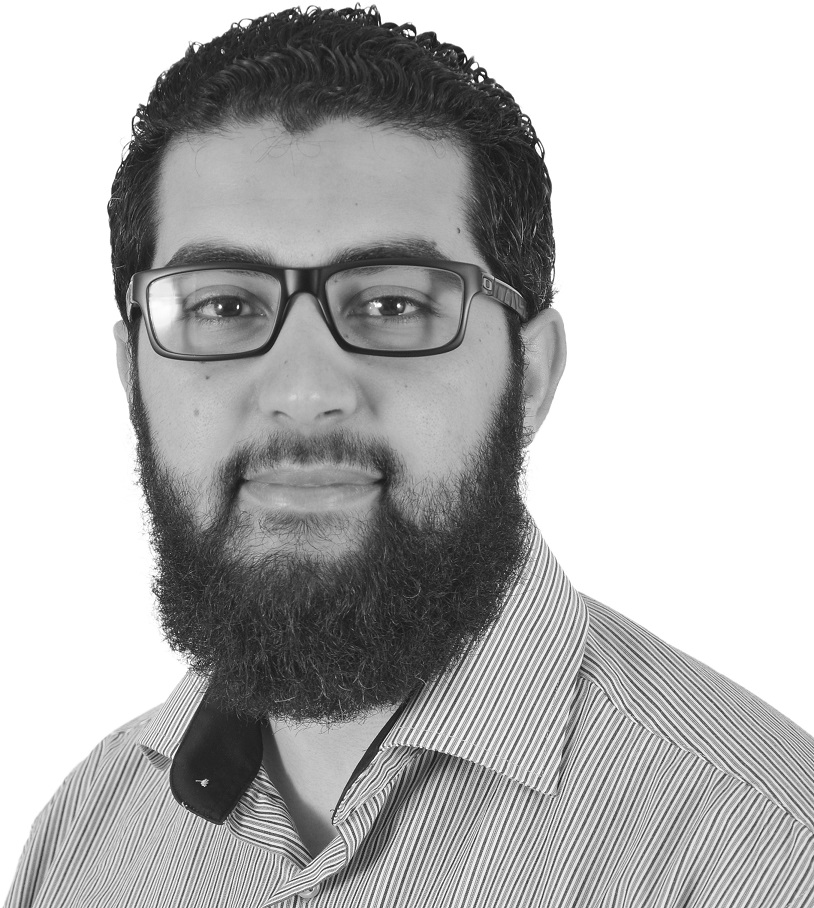}}]{Sid Ahmed Fezza}
received his engineer degree from University of Dr. Tahar MOULAY, Saïda, Algeria, in 2007 and the Ph.D. degree from Djillali Liabes University of Sidi-Bel-abbes, Algeria, in 2015, both in computer science. He is currently Associate Professor at National Institute of Telecommunications and ICT (INTTIC), Oran, Algeria. Sid Ahmed Fezza was a recipient of two top 10\% best paper awards in ICIP 2014, the 2015 Algerian Paper of the Year Awards from the Algerian Network for Academics, Scientists and Researchers and has authored several publications in top journals and conferences on image and video processing. His research interests are in the fields of image/video processing, image/video coding, visual quality assessment, immersive multimedia communication and multimedia security.
\end{IEEEbiography}

\begin{IEEEbiography}[{\includegraphics[width=1in,height=1.25in,clip,keepaspectratio]{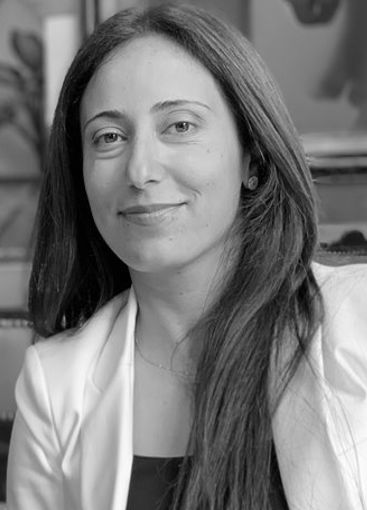}}]{Khouloud Samrouth}
received the master’s degree in informatics and telecommunication engineering from the Department of Electronics and Telecommunication, Faculty of Engineering, Lebanese University, Tripoli, Lebanon, in 2011, and the Ph.D. degree in signal and image processing from Lebanese University and National Institute of Applied Sciences of Rennes, Institute of Electronics and Telecommunications of Rennes, Rennes, France, in 2014. Currently, she is an assistant professor and a researcher at the Faculty of Engineering, Electronic and Telecommunication Department at Lebanese University. Her research is in the image processing field.
\end{IEEEbiography}
\begin{IEEEbiography}[{\includegraphics[width=1in,height=1.25in,clip,keepaspectratio]{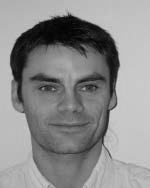}}]{Olivier D\'eforges}
received the Ph.D. degree in image processing in 1995. He is a Professor with the National Institute of Applied Sciences (INSA) of Rennes. In 1996, he joined the Department of Electronic Engineering, INSA of Rennes, Scientic and Technical University. He is a member of the Institute of Electronics and Telecommunications of Rennes (IETR), UMR CNRS 6164 and leads the IMAGE Team, IETR Laboratory including 40 researchers. He has authored over 130 technical papers. His principal research interests are image and video lossy and lossless compression, image understanding, fast prototyping, and parallel architectures. He has also been involved in the ISO/MPEG standardization group since 2007.
\end{IEEEbiography}

\end{document}